\documentclass[fleqn,usenatbib]{mnras}

\usepackage{newtxtext,newtxmath}

\usepackage[T1]{fontenc}

\DeclareRobustCommand{\VAN}[3]{#2}
\let\VANthebibliography\thebibliography
\def\thebibliography{\DeclareRobustCommand{\VAN}[3]{##3}\VANthebibliography}

\usepackage{graphicx}	
\usepackage{amsmath}	

\title[Comparing transit spectroscopy pipelines]{Comparing transit spectroscopy pipelines at the catalogue level: evidence for systematic differences}

\author[Mugnai L. V. et al.]{
Lorenzo V. Mugnai$^{1,2,3,4}$,\thanks{E-mail: mugnail@cardiff.ac.uk}
Mark R. Swain$^{5}$,
Raissa Estrela$^{5}$,
Gael M. Roudier$^{5}$
\\
$^{1}$School of Physics and Astronomy, Cardiff University, Queens Buildings, The Parade, Cardiff, CF24 3AA, UK\\
$^{2}$Dipartimento di Fisica, La Sapienza Universit\`a di Roma, Piazzale Aldo Moro 2, 00185 Roma, Italy\\
$^{3}$INAF – Osservatorio Astronomico di Palermo, Piazza del Parlamento 1, I-90134 Palermo, Italy\\
$^{4}$Department of Physics and Astronomy, University College London, Gower Street, London, WC1E 6BT, UK\\
$^{5}$Jet Propulsion Laboratory, California Institute of Technology, 4800 Oak Grove Drive, Pasadena, CA 91109, USA
}

\date{Accepted XXX. Received YYY; in original form ZZZ}

\pubyear{2023}

\begin{document}

\label{firstpage}
\pagerange{\pageref{firstpage}--\pageref{lastpage}}
\maketitle

\begin{abstract}
The challenge of inconsistent results from different data pipelines, even when starting from identical data, is a recognized concern in exoplanetary science. As we transition into the James Webb Space Telescope (JWST) era and prepare for the ARIEL space mission, addressing this issue becomes paramount because of its implications on our understanding of exoplanets. Although comparing pipeline results for individual exoplanets has become more common, this study is the first to compare pipeline results at the catalogue level. We present a comprehensive framework to statistically compare the outcomes of data analysis reduction on a population of exoplanets and we leverage the large number of observations conducted using the same instrument configured with HST-WFC3. We employ three independent pipelines: Iraclis, EXCALIBUR, and CASCADe. Our combined findings reveal that these pipelines, despite starting from the same data and planet system parameters, yield substantially different spectra in some cases. However, the most significant manifestations of pipeline differences are observed in the compositional trends of the resulting exoplanet catalogues. We conclude that pipeline-induced differences lead to biases in the retrieved information, which are not reflected in the retrieved uncertainties. Our findings underscore the critical need to confront these pipeline differences to ensure the reproducibility, accuracy, and reliability of results in exoplanetary research. Our results demonstrate the need to understand the potential for population-level bias that pipelines may inject, which could compromise our understanding of exoplanets as a class of objects.
\end{abstract}




\section{Introduction} \label{sec:intro}

A successful methodology for detecting atomic and molecular species and unveiling the atmospheric chemistry of exoplanets involves the use of multi-band transit photometry and spectroscopy \citep[e.g.][]{Charbonneau2002, Tinetti2007, Swain2008, Swain2009, Tsiaras2016b, Chachan2019, Mugnai2021b, Swain2021}. Current space instrumentation, such as the Spitzer and Hubble Space Telescopes, have facilitated the atmospheric characterization of approximately sixty exoplanets over a limited wavelength range \citep[e.g.][]{Sing2016, iyer2016, Barstow2017, Tsiaras2018, Edwards2022, 2022Estrela}. However, these instruments were not specifically designed for exoplanetary science, necessitating specialized data reduction pipelines to remove instrument systematics that are similar in amplitude to the astrophysical signal \citep{Deming2013, Tsiaras2016}.

To interpret the observed spectra, spectral retrieval techniques are commonly used to estimate astrophysical parameters  \citep[e.g.][]{Irwin2008, Madhusudhan2009, Line2013, Lee2013, Waldmann2015, GandhiMadhusudhan2017, Lavie2017, AlRefaie2021}. Studies have been conducted to compare and validate different retrieval models, demonstrating their robustness and consistency \citep{Barstow2020, Barstow2022}. However, a similar in-depth large-scale validation has not been performed for data reduction pipelines, which estimate the spectra from raw data. Uncharacterized biases introduced at this stage of data analysis can potentially undermine the correct interpretation of observations using retrieval techniques. While the recent literature does chronicle multiple validation endeavours, these comparisons are undertaken on a singular-planet basis. A remarkable example of this trend is offered by the Early Release Science of James Webb Telescope: \cite{2023G395H, 2023PRISM, 2023NIRCam, 2023NIRISS,Madhusudhan2023}. Thus, there is a compelling need for holistic, population-centric validation, which is the cornerstone of our proposed study.

The data reduction process for exoplanet transit spectroscopy has a number of steps where differences in methods have the potential to produce differences in final outcomes. A nonexhaustive list of specific areas where method differences might influence the final outcome includes; spectral extraction and background subtraction, interpolation errors associated with placing spectra on a common wavelength grid, system parameter values, astrophysical models such as the detailed formulation of the limb-darkening relation, outlier rejection methods, the value or width of any priors applied, which parameters are locked and which are retrieved, the dimensionality and form of the instrument model, and the formulation of the sampler. Differences in the method can lead to differences in astrophysical interpretation \citep[e.g.][]{Swain2021,Mugnai2021b, Libby-Roberts2022} and raise the question of which result more accurately represents the astrophysical reality. Further reinforcing the notion that differences in methods can impact the astrophysical interpretation, the literature is replete with examples of exoplanet descriptions being revised due to different approaches to the modelling and removal of systematics in data reduction pipelines \citep[e.g.][]{DiamondLowe2014, Stevenson2014a, Stevenson2014b, Tsiaras2018}. A notable example is the hypothesis of six exomoon candidates proposed by \citet{FoxWiegert2020}, which was later discarded by an updated data reduction pipeline \citep{Kipping2020}. Using different stellar or planetary parameters for the analysis is known to introduce offsets or slopes in the dataset \citep{Morello2017, Alexoudi2018}. Finally, time variability can arise from stellar activity \citep{Bruno2020, Kirk2019}.

A further complication is the potential inconsistency between datasets from different instruments, which remains a problematic issue that has been tentatively discussed in the literature to assess its implications for our understanding of exoplanets. \citep[e.g.][]{Yip2020, Pluriel2020, Saba2022}. Different data analysis approaches have been suggested to take into account these discrepancies during retrieval \citep{Yip2021}, however, this issue warrants further research to ensure the reliability and reproducibility of exoplanetary science. In particular, with the anticipated contributions from the James Webb Space Telescope (JWST), ensuring consistency across multi-instrument datasets will be paramount for the accurate interpretation and understanding of exoplanetary atmospheres \citep{Constantinou2023}.

In fact, as we usher in the era of the James Webb Space Telescope (JWST) and the ARIEL space mission, the challenges posed by the possibility of pipeline-dependent biases operating on an entire observational catalogue must be taken seriously and understood. The advent of these next-generation telescopes promises unprecedented data quality, allowing researchers to delve into more intricate questions about exoplanets. However, the very richness of this data also amplifies the potential pitfalls of pipeline discrepancies. While this study highlights issues observed with HST-WFC3 data, the implications extend far beyond. In the JWST and ARIEL era, where the focus will be on planetary population studies and comparative planetology, ensuring consistency and reliability across pipelines is paramount. Addressing these discrepancies is not just about refining our current understanding of exoplanets but is crucial for harnessing the full potential of upcoming observational capabilities. Only by resolving the challenge posed by pipeline-dependent results can we truly capitalize on the advanced data, asking deeper questions and drawing more precise conclusions about the universe's myriad exoplanets.



In this study, we aim to explore the existence of systematic biases originating from the analysis processes of different pipelines in the examination of exoplanet catalogues. It's crucial to note that our objective is not to pinpoint the precise origins of differences between pipelines, a task that would necessitate a detailed comparison of standardized intermediate data products to locate where discrepancies are injected.

\section{Methods}

\subsection{The datasets} \label{sec:dataset}
\begin{figure}
    \centering
    \includegraphics[width=\columnwidth]{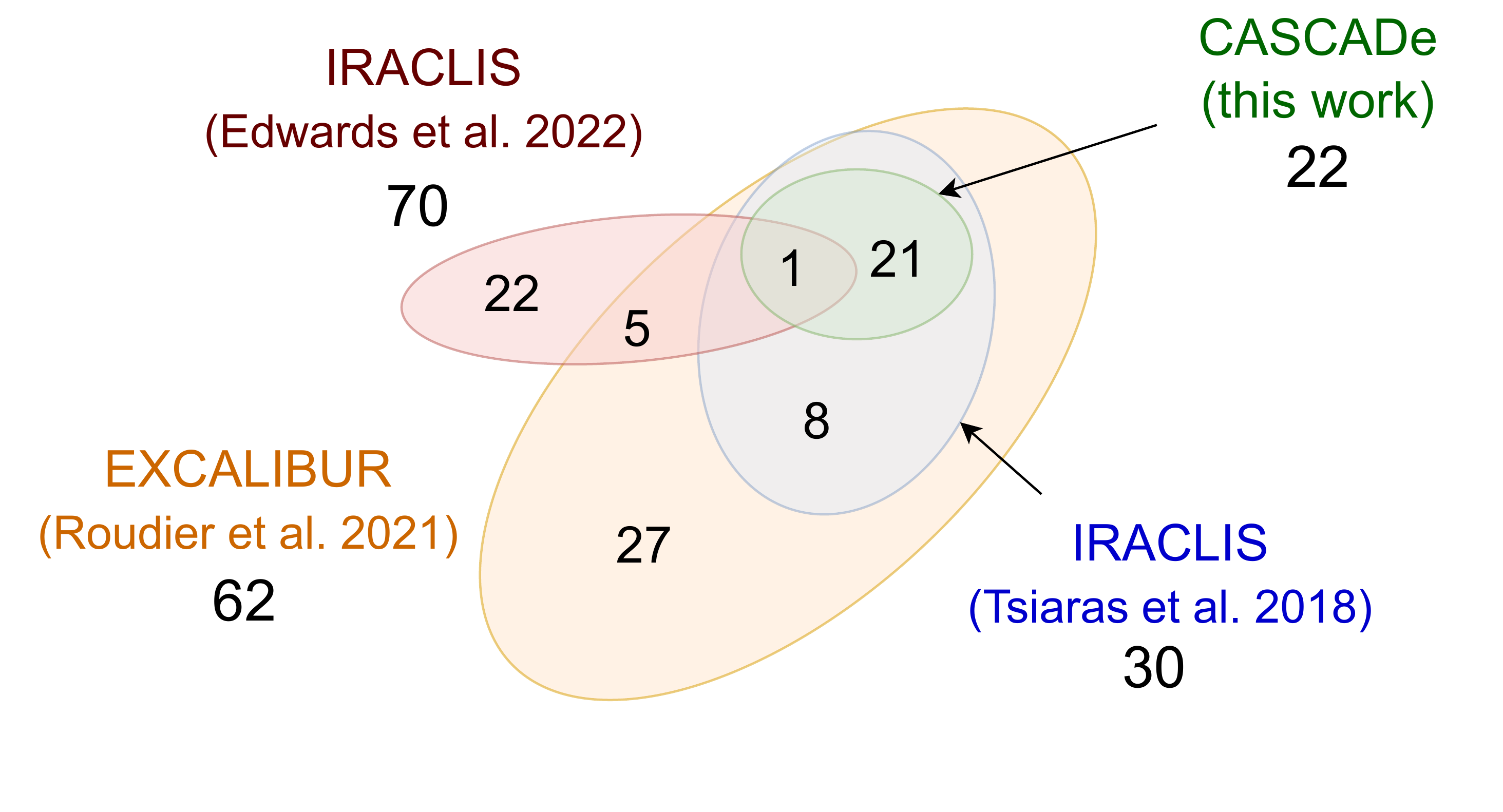}
    \caption{This study integrates four datasets of transmission spectra, derived from HST-WFC3 observations, as shown on the left. The first dataset (blue), based on the Iraclis pipeline, is adopted from \citet{Tsiaras2018} and encompasses 30 transmission spectra. Also based on the Iraclis pipeline, a second dataset (red) was curated and published in \citet{Edwards2022}, contributing 6 planetary spectra to this study. The reader should note here that the sum in the red ellipse gives 28 and not 70. This is because here we are considering only the planets consistently analyzed in \citet{Edwards2022}, which are 28, and not the planetary spectra that the authors used in their work, but were processed somewhere else. A third dataset (orange), constructed with the EXCALIBUR pipeline, is outlined in \citet{Roudier2021} and contains 62 spectra. Of these, 30 correspond to planets analyzed in \citet{Tsiaras2018} and overlap with that dataset, and 6 overlap with the planets from \citet{Edwards2022}. The fourth dataset (green), developed through the CASCADe pipeline for this work, employs the planetary parameters from \citet{Roudier2021} and reported in Tab. \ref{tab:star_pars} and \ref{tab:pln_pars}. This novel dataset includes 22 spectra that coincide with the other datasets. In this case, we consider only 22 planets and not 30, because for 8 planets the automatic pipeline failed for different reasons: because the goal of this work is to compare consistently analyzed populations, we decided to exclude the 8 planets instead of proceeding with dedicated data processing. The figure highlights the intersections between datasets and enumerates the spectra contained in each intersection. Note that one planet, WASP-121b, is shared between all the datasets.}
    \label{fig:database}
\end{figure}

In this study, we compare the spectra produced starting from three pipelines (Iraclis, EXCALIBUR and CASCADe) and four different datasets. A summary representation is reported in Fig. \ref{fig:database}.  

\begin{figure*}
    \centering
    \includegraphics[width=0.8\textwidth]{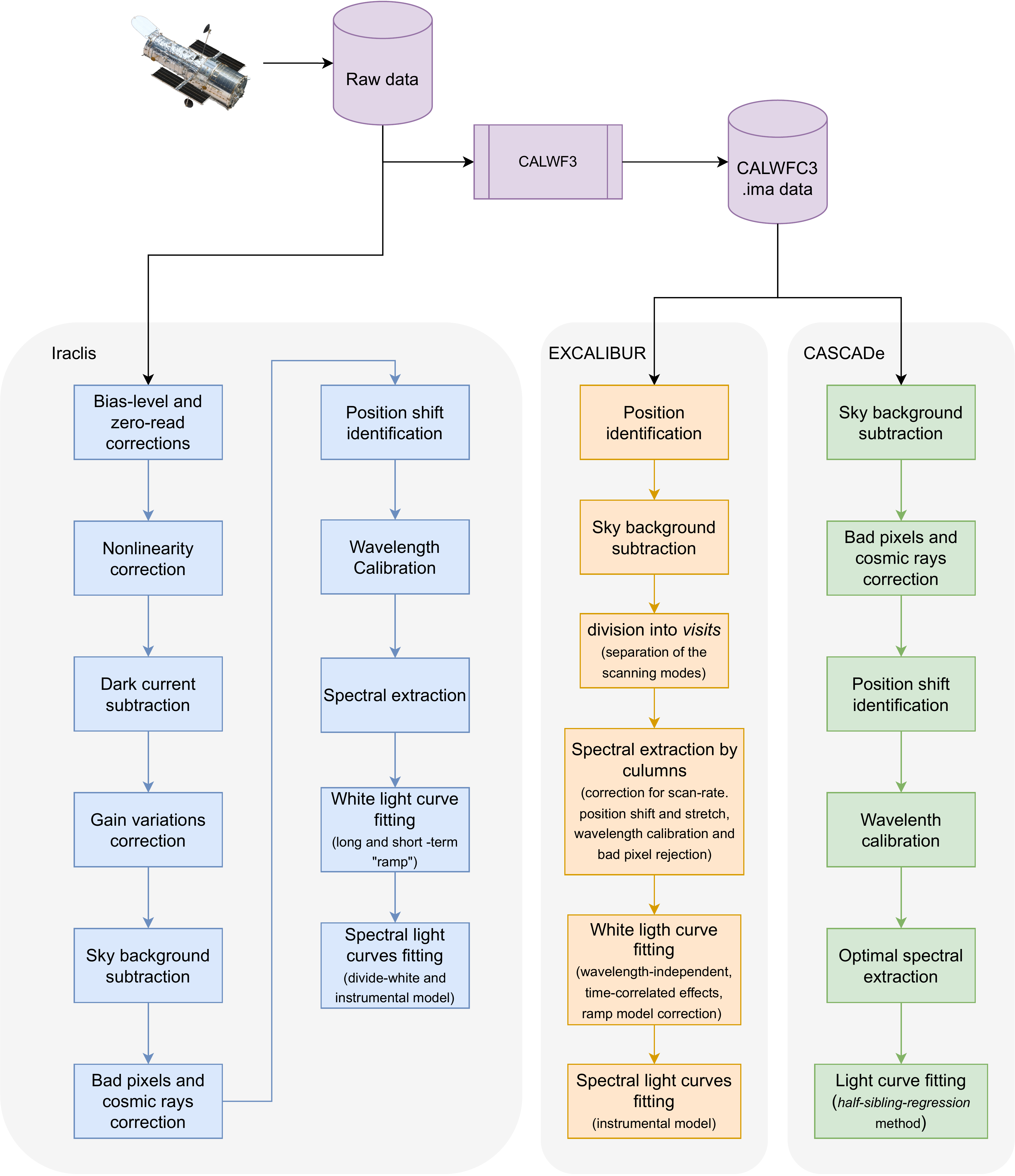}
    \caption{Comparison of data reduction steps across pipelines. The data reduction steps for the Iraclis pipeline are delineated following \citet{Tsiaras2016}, for EXCALIBUR as per \citet{Swain2021}, and for CASCADe within the Appendices of \citet{Carone2021}. These pipelines, while employing conceptually similar strategies for data processing, differ significantly in the specifics of their implementation: examples of these different implementations are reported between parenthesis in the diagram. An in-depth analysis of these differences falls outside the scope of this paper, and readers are referred to the respective primary sources for detailed methodologies.}
    \label{fig:pipesteps}
\end{figure*}

\subsubsection{Iraclis: Tsiaras et al. 2018}
The Iraclis database, as presented by \citet{Tsiaras2018}, contains transmission spectra of 30 gaseous planets, generated using the Iraclis pipeline. This pipeline executes a series of steps, including zero-read subtraction, reference pixel correction, nonlinearity correction, dark current subtraction, gain conversion, sky-background subtraction, calibration, flatfield correction, and bad pixels and cosmic-ray correction. The pipeline then extracts the flux from the spatially scanned spectroscopic images to produce the final transit light curves per wavelength band. The light curves are fitted using literature values, with the planet-to-star radius ratio and the transit midtime as the only free parameters, apart from the coefficients for Hubble systematics. The limb-darkening coefficients are selected from the quadratic formula by \citet{Claret2000}, using the stellar parameters. These models have been recently incorporated into the ExoTethys package \citep{Morello2020} utilized by Iraclis. The spectral light curves are then fitted using the divide-white method introduced by \citet{Kreidberg2014}, with the inclusion of a normalization factor for the slope. A summary of the data reduction steps is reported in Fig. \ref{fig:pipesteps}

\subsubsection{Iraclis: Edwards et al. 2022}
A subsequent dataset, also processed using the Iraclis pipeline, was introduced by \citet{Edwards2022}. This dataset contains 70 transmission spectra of gaseous planets processed with the Iraclis pipeline, with 29 derived from \citet{Tsiaras2018} and 13 obtained from other sources such as \citet{Anisman2020, Changeat2020, Edwards2020, Guilluy2021, skaf2020, Pluriel2020, Saba2022, Tsiaras2019, Yip2021}. The present study focuses exclusively on the latest 28 transmission spectra, as they were specifically processed for the work presented in \citet{Edwards2022} using an updated version of the software. Of these spectra, only 6 were used because they coincide with spectra contained in other datasets. One of the six planets is also shared with the \citet{Tsiaras2018} paper: WASP-121b. A deeper discussion on this planet is reported in sec. \ref{sec:wasp-121}.

\subsubsection{EXCALIBUR: Roudier et al. 2021}
The EXCALIBUR 2021 catalogue, presented by \citet{Roudier2021}, contains 62  transmission spectra obtained using the EXCALIBUR pipeline\footnote{The data are available at \url{http://excalibur.ipac.caltech.edu/}, referring to EXCALIBUR Run ID \texttt{187}.}. EXCALIBUR uses a fully automated, uniform processing approach with persistent intermediate data products to maintain the chain of inference. Further information EXCALIBUR can be found in \citet{Swain2021,Roudier2021, HuberFeely2022} and a scheme of the pipeline steps is reported in Fig. \ref{fig:pipesteps}. Of the 62 planets in the EXCALIBUR 2021 catalogue, 30 spectra correspond to planets also processed in \citet{Tsiaras2018}. These planets are XO-1b, WASP-31b, HAT-P-38b, HAT-P-41b, HD 209458b, WASP-69b, WASP-76b, WASP-121b, WASP-43b, WASP-52b, WASP-74b, WASP-101b, HD 149026b, HAT-P-17b, HAT-P-12b, WASP-63b, HD 189733b, HAT-P-26b, GJ 436b, HAT-P-11b, HAT-P-32b, WASP-67b, WASP-39b, WASP-80b, GJ 3470b, WASP-29b, WASP-12b, HAT-P-3b, HAT-P-18b, and HAT-P-1b. Additionally, the Excalibur dataset shares six planets with the dataset presented in \citet{Edwards2022}: WASP-121b, WASP-107b, GJ 1214b, KELT-1b, K2-24b, and WASP-18b.

\subsubsection{CASCADe: this work}
In this work, we also constructed a database using an automated procedure within the CASCADe\footnote{For this work we use the code version \texttt{1.1.15}, which is available at \url{ https://jbouwman.gitlab.io/CASCADe/}.} pipeline, maintaining the same planetary parameters utilized in \citet{Roudier2021} and listed later in the text in Tab. \ref{tab:star_pars} and \ref{tab:pln_pars}. CASCADe represents an instrument-independent reduction pipeline that has demonstrated its versatility through application to data sets from both the Hubble Space Telescope (HST) and the Spitzer Space Telescope \citep{2020ASPC..527..179L, Carone2021}. Furthermore, its efficacy has been validated through tests on simulations from the James Webb Space Telescope's Mid-Infrared Instrument (JWST MIRI). Similar to EXCALIBUR, the CASCADe pipeline initiates the data reduction process with the ``ima'' intermediate data product. This product is generated by the CALWFC3 data reduction pipeline, marking a departure from pipelines such as Iraclis, which undertake the data calibration themselves. A novel feature of CASCADe is its implementation of a data-driven method (the \textit{half-sibling-regression} method), a pioneering approach introduced by \citet{2016PNAS..113.7391S}. This method leverages the causal connections inherent within a data set to calibrate the spectral time-series data, potentially enhancing the accuracy and reliability of the data reduction process. For a full description of CASCADe, refer to \citet{Carone2021}, while Fig. \ref{fig:pipesteps} reports a summary of the data reduction steps as reported in that work appendices.
This dataset comprises 22 transmission spectra, which are also found in \citet{Tsiaras2018}: WASP-31b, HAT-P-41b, WASP-76b, WASP-121b, which is also found in \citet{Edwards2022}, WASP-52b, WASP-74b, HD 149026b, HAT-P-17b, HAT-P-12b, WASP-63b, HAT-P-26b, GJ 436b, HAT-P-11b, HAT-P-32b, WASP-67b, WASP-39b, WASP-80b, GJ 3470b, WASP-29b, HAT-P-3b, HAT-P-18b, and HAT-P-1b. All of these spectra overlap with the EXCALIBUR data processed in \citet{Roudier2021}. We present in this work only 22 spectra produced with the CASCADe pipeline, instead of fully reproducing the other datasets because we decided to focus on the planets presented in \citet{Tsiaras2018}, as these spectra are publicly available and the reader can reproduce our results. Moreover, we decided to run the CASCADe pipeline fully automatically, with the only exception of the control on the stellar and planetary parameter adapted from \citet{Roudier2021}. So, we decided to exclude every planetary observation that requires custom analysis.

\subsubsection{Catalogues summary}
We included 35 planets in our study. The three datasets share 22 planets: WASP-67b, WASP-31b, HD 149026b, HAT-P-41b, HAT-P-1b, WASP-76b, WASP-74b, HAT-P-12b, HAT-P-17b, HAT-P-26b, WASP-39b, WASP-52b, GJ 436b, HAT-P-32b, HAT-P-11b, HAT-P-3b, WASP-63b, HAT-P-18b, WASP-121b, WASP-80b, GJ 3470b, and WASP-29b. Also, all the datasets include WASP-121b: a detailed discussion around this planet can be found in sec. \ref{sec:wasp-121}. We aim to leverage these intersections between the datasets to infer pipeline information on a statistical basis.
Such intersections are highlighted in Fig. \ref{fig:database}. To help with the identification of the database in the following we use the notation
\begin{itemize}
  \setlength\itemsep{0em}
    \item ``Iraclis'' for the database presented in \citet{Tsiaras2018};
    \item ``Edwards2022'' for the database presented in \citet{Edwards2022};
    \item ``Excalibur'' for the database presented in \citet{Roudier2021};
    \item ``Cascade'' for the database produced for this work with the CASCADe pipeline.
\end{itemize}

\begin{table}
    \centering
    \begin{tabular}{|c|c|c|c|c|}
        \hline
         Catalogue& Iraclis & Edwards2022 & Excalibur & Cascade\\
         {\scriptsize Reference} &  {\scriptsize TS18}& {\scriptsize ED22} & {\scriptsize RO21} & {\scriptsize This Work} \\
         {\scriptsize Pipeline}& {\scriptsize Iraclis} & {\scriptsize Iraclis} & {\scriptsize EXCALIBUR} & {\scriptsize CASCADe} \\
         \hline
    GJ 1214b & - - & \checkmark & \checkmark &  - -\\
    GJ 3470b & \checkmark & - - & \checkmark & \checkmark \\ 
    GJ 436b & \checkmark &  - -& \checkmark & \checkmark \\ 
    HAT-P-11b & \checkmark & - - & \checkmark & \checkmark \\ 
    HAT-P-12b & \checkmark & - - & \checkmark & \checkmark \\ 
    HAT-P-17b & \checkmark & - - & \checkmark & \checkmark \\ 
    HAT-P-18b & \checkmark & - - & \checkmark & \checkmark \\ 
    HAT-P-1b & \checkmark & - - & \checkmark & \checkmark \\ 
    HAT-P-26b & \checkmark & - - & \checkmark & \checkmark \\ 
    HAT-P-32b & \checkmark & - - & \checkmark & \checkmark \\ 
    HAT-P-38b & \checkmark & - - & \checkmark &  - -\\ 
    HAT-P-3b & \checkmark & - - & \checkmark & \checkmark \\ 
    HAT-P-41b & \checkmark & - - & \checkmark & \checkmark \\ 
    HD 149026b & \checkmark & - - & \checkmark & \checkmark \\ 
    HD 189733b & \checkmark & - - & \checkmark &  - -\\
    HD 209458b & \checkmark & - - & \checkmark & - - \\ 
    K2-24b & - - & \checkmark & \checkmark &  - -\\ 
    KELT-1b & - - & \checkmark & \checkmark & - - \\ 
    WASP-101b & \checkmark & - - & \checkmark & - - \\ 
    WASP-107b & - - & \checkmark & \checkmark & \checkmark \\ 
    WASP-121b & \checkmark & \checkmark & \checkmark & \checkmark \\
    WASP-12b & \checkmark & - - & \checkmark & - - \\
    WASP-18b & - - & \checkmark & \checkmark & - - \\
    WASP-29b & \checkmark & - - & \checkmark & \checkmark \\ 
    WASP-31b & \checkmark & - - & \checkmark & \checkmark \\ 
    WASP-39b & \checkmark & - - & \checkmark & \checkmark \\ 
    WASP-43b & \checkmark & - - & \checkmark & - -\\ 
    WASP-52b & \checkmark & - - & \checkmark & \checkmark \\ 
    WASP-63b & \checkmark & - - & \checkmark & \checkmark \\ 
    WASP-67b & \checkmark & - - & \checkmark & \checkmark \\ 
    WASP-69b & \checkmark & - - & \checkmark &  - -\\ 
    WASP-74b & \checkmark & - - & \checkmark & \checkmark \\
    WASP-76b & \checkmark & - - & \checkmark & \checkmark \\ 
    WASP-80b & \checkmark & - - & \checkmark & \checkmark \\ 
    XO-1b & \checkmark & - - & \checkmark &  - -\\ 

        \hline 
    \end{tabular}
    \caption{List of all the planets included in this work and of the datasets containing their transmission spectra. The header includes the catalogue name (first row), the catalogue reference paper (second row), and the pipeline used (third row). TS18 refers to \citet{Tsiaras2018}, ED22 to \citet{Edwards2022} and RO21 to \citet{Roudier2021}.}
    \label{tab:planets_list}
\end{table}

\begin{table}
    \centering
\begin{tabular}{|l|c|c|c|c|} 
         \hline 
  Name &  $\textrm{log} \, g_{\star}$ &  $\textrm{T}_{\star}$ [$K$] &  $\textrm{R}_{\star}$ [$R_{\odot}$] &  [Fe/H]\\
\hline 
 GJ 1214 & 4.94 & 3026 & 0.22 & 0.39 \\ 
 GJ 3470 & 4.7 & 3600 & 0.55 & 0.20 \\ 
 GJ 436 & 4.79 & 3416 & 0.46 & 0.02 \\ 
 HAT-P-11 & 4.66 & 4780 & 0.68 & 0.31 \\ 
 HAT-P-12 & 4.61 & 4650 & 0.70 & -0.29 \\ 
 HAT-P-17 & 4.53 & 5246 & 0.84 & 0.00 \\ 
 HAT-P-18 & 4.57 & 4803 & 0.75 & 0.10 \\ 
 HAT-P-1 & 4.36 & 5980 & 1.17 & 0.13 \\ 
 HAT-P-26 & 4.56 & 5079 & 0.87 & -0.04 \\ 
 HAT-P-32 & 4.22 & 6001 & 1.37 & -0.16 \\ 
 HAT-P-38 & 4.45 & 5330 & 0.92 & 0.06 \\ 
 HAT-P-3 & 4.56 & 5185 & 0.87 & 0.27 \\ 
 HAT-P-41 & 4.14 & 6390 & 1.68 & 0.21 \\ 
 HD 149026 & 4.37 & 6179 & 1.41 & 0.32 \\ 
 HD 189733 & 4.49 & 5052 & 0.75 & -0.02 \\ 
 HD 209458 & 4.45 & 6091 & 1.19 & 0.01 \\ 
 K2-24 & 4.29 & 5625 & 1.16 & 0.34 \\ 
 KELT-1 & 4.228 & 6518 & 1.46 & 0.01 \\ 
 WASP-101 & 4.31 & 6380 & 1.31 & 0.20 \\ 
 WASP-107 & 4.5 & 4430 & 0.66 & 0.02 \\ 
 WASP-121 & 4.24 & 6459 & 1.46 & 0.13 \\ 
 WASP-12 & 4.38 & 6300 & 1.59 & 0.30 \\ 
 WASP-18 & 4.47 & 6431 & 1.29 & 0.13 \\ 
 WASP-29 & 4.5 & 4800 & 0.79 & 0.11 \\ 
 WASP-31 & 4.76 & 6302 & 1.25 & -0.08 \\ 
 WASP-39 & 4.48 & 5400 & 0.90 & -0.10 \\ 
 WASP-43 & 4.646 & 4400 & 0.60 & -0.05 \\ 
 WASP-52 & 4.58 & 5000 & 0.79 & 0.03 \\ 
 WASP-63 & 4.01 & 5550 & 1.86 & 0.08 \\ 
 WASP-67 & 4.35 & 5200 & 0.88 & -0.07 \\ 
 WASP-69 & 4.5 & 4700 & 0.86 & 0.15 \\ 
 WASP-74 & 4.39 & 5990 & 1.42 & 0.39 \\ 
 WASP-76 & 4.13 & 6250 & 1.73 & 0.23 \\ 
 WASP-80 & 4.66 & 4143 & 0.59 & -0.13 \\ 
 XO-1 & 4.51 & 5750 & 0.88 & 0.02 \\ 
\hline 
 
 \end{tabular}
  \caption{List of stellar parameters used in this study. All the parameters are from \citet{Roudier2021} and listed in the EXCALIBUR archive. The reader can refer to that work and the reference therein for further details. For HAT-P-38, KELT-1, and WASP-43, the $\textrm{log} \, g_{\star}$ values were estimated by the EXCALIBUR pipeline due to the absence of direct measurements in the literature.}
    \label{tab:star_pars}
\end{table}

\begin{table*}
    \centering
\begin{tabular}{|l|c|c|c|c|c|c|c|c|} 
         \hline 
Name & Period [\textit{days}] & $\textrm{T}_{\textrm{eq}}$ [$K$] & a [$au$] & R [$R_{Jup}$] & M [$M_{Jup}$] & Inc. [$degree$] & Ecc. & t$_0$ [\textit{Julian Days}] \\ 
\hline 
GJ 1214b & 1.58040456 & 561.2 & 0.0141 & 0.254 & 0.02 & 88.17 & 0.0 & 2455320.535733 \\ 
GJ 3470b & 3.3366496 & 665.5 & 0.0355 & 0.408 & 0.044 & 89.13 & 0.017 & 2455983.70421 \\ 
GJ 436b & 2.64389782 & 619.8 & 0.0308 & 0.372 & 0.07 & 86.774 & 0.0 & 2456295.431924 \\ 
HAT-P-11b & 4.887802443 & 809.4 & 0.0525 & 0.389 & 0.074 & 90.0 & 0.0 & 2454957.8132067 \\ 
HAT-P-12b & 3.2130598 & 932.5 & 0.0384 & 0.959 & 0.211 & 89.0 & 0.0 & 2454419.19556 \\ 
HAT-P-17b & 10.338523 & 920.8 & 0.06 & 1.05 & 0.58 & 89.2 & 0.35 & 2454801.16945 \\ 
HAT-P-18b & 5.508023 & 826.3 & 0.0559 & 0.995 & 0.197 & 88.8 & 0.084 & 2454715.02174 \\ 
HAT-P-1b & 4.46529976 & 1288.3 & 0.0556 & 1.319 & 0.525 & 85.634 & 0.0 & 2453979.92802 \\ 
HAT-P-26b & 4.23452 & 1016.6 & 0.0479 & 0.63 & 0.07 & 88.6 & 0.12 & 2455304.6522 \\ 
HAT-P-32b & 2.1500082 & 1789.9 & 0.034 & 1.98 & 0.68 & 88.98 & 0.159 & 2455867.402743 \\ 
HAT-P-38b & 4.640382 & 1049.9 & 0.0523 & 0.825 & 0.267 & 88.3 & 0.067 & 2455863.11957 \\ 
HAT-P-3b & 2.8997 & 1151.1 & 0.0389 & 0.94 & 0.65 & 87.24 & 0.0 & 2454218.81 \\ 
HAT-P-41b & 2.694047 & 1886.4 & 0.0426 & 2.05 & 1.19 & 87.7 & 0.0 & 2454983.86167 \\ 
HD 149026b & 2.87589 & 1649.6 & 0.0436 & 0.74 & 0.38 & 84.55 & 0.0 & 2454597.7071 \\ 
HD 189733b & 2.21857567 & 1161.5 & 0.0313 & 1.13 & 1.13 & 85.71 & 0.0 & 2453955.5256 \\ 
HD 209458b & 3.52474859 & 1438.4 & 0.0471 & 1.39 & 0.73 & 86.71 & 0.0 & 2452826.6293 \\ 
K2-24b & 20.88977 & 725.1 & 0.154 & 0.482 & 0.06 & 88.874 & 0.06 & 2456905.8855 \\ 
KELT-1b & 1.217513 & 2353.6 & 0.0247 & 1.11 & 27.2 & 87.6 & 0.0 & 2455933.61 \\ 
WASP-101b & 3.58572 & 1525.1 & 0.0506 & 1.43 & 0.51 & 85.0 & 0.0 & 2456164.6941 \\ 
WASP-107b & 5.72149 & 720.8 & 0.055 & 0.94 & 0.12 & 89.7 & 0.0 & 2456514.4106 \\ 
WASP-121b & 1.2749255 & 2298.1 & 0.0254 & 1.865 & 1.18 & 87.6 & 0.0 & 2456635.70832 \\ 
WASP-12b & 1.09142245 & 2439.1 & 0.0234 & 1.937 & 1.46 & 82.5 & 0.0447 & 2456176.6683 \\ 
WASP-18b & 0.94145 & 2413.7 & 0.0202 & 1.2 & 11.4 & 80.6 & 0.01 & 2455265.5525 \\ 
WASP-29b & 3.92273 & 937.3 & 0.0457 & 0.77 & 0.23 & 88.8 & 0.03 & 2455830.1889 \\ 
WASP-31b & 3.4059096 & 1533.1 & 0.0466 & 1.549 & 0.478 & 84.41 & 0.0 & 2455192.6887 \\ 
WASP-39b & 4.055259 & 1091.4 & 0.0486 & 1.27 & 0.28 & 87.83 & 0.0 & 2455342.9688 \\ 
WASP-43b & 0.813475 & 1343.3 & 0.0142 & 0.93 & 1.78 & 82.6 & 0.0 & 2455528.86774 \\ 
WASP-52b & 1.7497798 & 1265.6 & 0.0272 & 1.27 & 0.46 & 85.35 & 0.0 & 2455793.68143 \\ 
WASP-63b & 4.37808 & 1483.8 & 0.0574 & 1.41 & 0.37 & 87.8 & 0.0 & 2455921.6536 \\ 
WASP-67b & 4.61442 & 1006.7 & 0.0518 & 1.15 & 0.43 & 85.8 & 0.0 & 2455824.375 \\ 
WASP-69b & 3.86814 & 962.3 & 0.0452 & 1.11 & 0.29 & 86.71 & 0.0 & 2455748.8342 \\ 
WASP-74b & 2.13775 & 1741.7 & 0.037 & 1.36 & 0.72 & 79.81 & 0.0 & 2456506.8926 \\ 
WASP-76b & 1.809886 & 2125.4 & 0.033 & 1.83 & 0.92 & 88.0 & 0.0 & 2456107.85507 \\ 
WASP-80b & 3.06785234 & 805.9 & 0.0344 & 0.999 & 0.538 & 89.02 & 0.002 & 2456487.425006 \\ 
XO-1b & 3.94153 & 1146.8 & 0.0488 & 1.14 & 0.83 & 88.81 & 0.0 & 2453887.7477 \\ 
\hline 
 
 \end{tabular}    
   \caption{List of planetary parameters used in this study. 
   All the parameters are from \citet{Roudier2021} and listed in the EXCALIBUR archive. The reader can refer to that work and to the reference therein for further details.}
    \label{tab:pln_pars}
\end{table*}

Also, a list of the planet spectra contained in each dataset is reported in Tab. \ref{tab:planets_list}, while lists of the parameters used from \citet{Roudier2021}\footnote{The data are also available with their references at \url{http://excalibur.ipac.caltech.edu/}, referring to EXCALIBUR Run ID \texttt{155}.} are reported in Tab. \ref{tab:star_pars} and \ref{tab:pln_pars}.

A schematic overview of the main steps involved in the three data processing pipelines is presented in Fig.~\ref{fig:pipesteps}. This illustration highlights the utilization of analogous methodologies for data detrending across the pipelines. Nonetheless, the specifics of the implementation for each pipeline vary considerably. An in-depth discussion of these implementation details falls beyond the scope of this article.

\subsection{Comparison strategy} \label{sec:strategy}

To enable a comparison of the spectra produced by various pipelines, we bin the spectra to match the spectral resolution used in the \citet{Tsiaras2018} dataset, which corresponds to a resolving power of 70 at $1.4 \, \mu m$. This procedure resulted in 25 data points for each spectrum. Consequently, the Excalibur dataset shows one or more empty bins for the following planets due to the EXCALIBUR outlier rejection approach \citep{Swain2021}: HAT-P-12b, HAT-P-32b, WASP-107b, WASP-31b, WASP-39b, WASP-43b.   

\begin{figure*}
    \centering
    \includegraphics[width=0.9\textwidth]{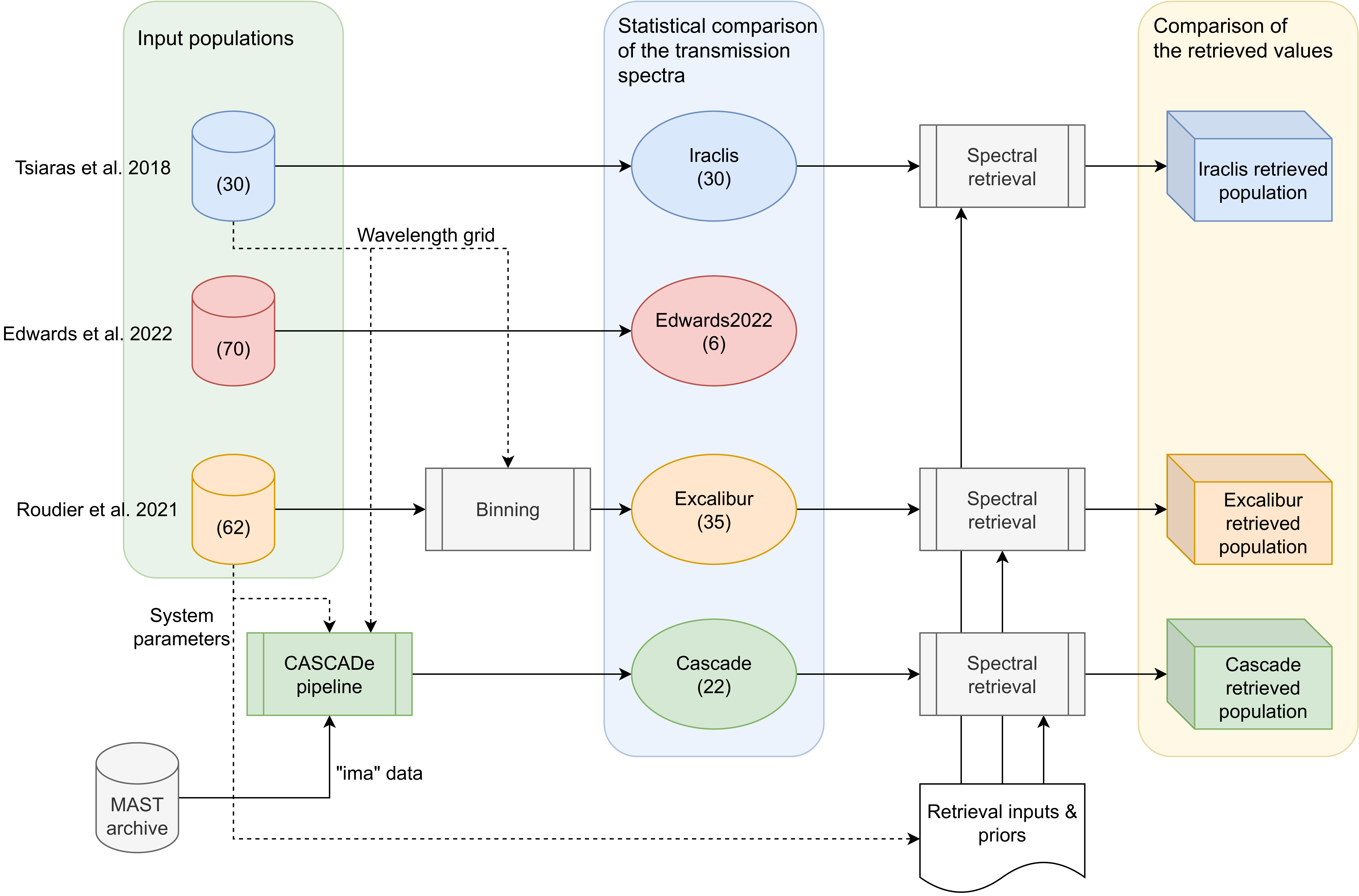}
    \caption{The figure summarizes the strategy used to compare the planetary transmission spectra. Starting from the input populations from published works, we binned down all the spectra to the same spectral resolution, and we used the same planetary parameters used in \citet{Roudier2021} to produce the Cascade dataset. We compare then the transmission spectra, obtaining the results reported in sec. \ref{sec:statistics}. Using the input parameters reported in Tab. \ref{tab:star_pars} and \ref{tab:pln_pars}, we perform spectral retrieval on each spectrum of the most populated datasets, using the same boundaries and priors. Finally, we compare the derived measurements in Sec. \ref{sec:retrieval}. }
    \label{fig:strategy}
\end{figure*}

Our analysis proceeded in two primary stages. Firstly, we compared the statistical properties of the spectra produced by different pipelines. The results of this comparison are reported in Sec. \ref{sec:statistics}. Secondly, to evaluate the implications of the observed differences on the retrieved information, we conducted consistent retrievals and compared the derived properties (Sec. \ref{sec:retrieval}). The described strategy, from the creation of the datasets, is summarised in Fig. \ref{fig:strategy}. As shown in the figure, we decided to not perform the retrieval study for the Edwards2022 catalogue. Further details behind this decision are reported in sec. \ref{sec:edwards}.

For the retrieval process, we utilized Alfnoor \citep{Alfnoor1, Mugnai2021}, a TauRex 3 \citep{taurex} wrapper that streamlines the retrieval procedure. We considered all planets to have primary atmospheres with He/H$_2$=0.17, with H$_2$O and CH$_4$ as trace gasses. The parallel plane approximation was adopted for an isothermal atmosphere, composed of 100 layers ranging from $10^{6}$ to $10^{-5}$ Pa. We also incorporated CIA \citep{abel_h2-h2,abel_h2-he, fletcher_h2-h2} and Rayleigh effects into the model. The star and planet parameters were sourced from \citet{Roudier2021}, and they are listed in Tab. \ref{tab:star_pars} and \ref{tab:pln_pars}. The planetary equilibrium temperature was computed for the temperature parameter as 
\begin{equation} \label{eq:temp}
    T_{eq} = T_{\star}\sqrt{\frac{R_{\star}}{2 \, a}}(1-A)^{\frac{1}{4}}    
\end{equation}
where $T_{\star}$ is the stellar temperature, $R_{\star}$ the stellar radius, $a$ is the planet's semimajor axis and $A$ is the Bond albedo that we arbitrarily set to 0.1 for all the planets.

In the fitting procedure, we investigate the planet radius with uniform priors between $0.1$ and $10$ times the input value, the temperature with uniform priors between $0.5$ and $1.5$ times the equilibrium temperature, a cloud deck (represented as grey clouds, using logarithmic uniform priors from $10^{6}$ to $10^{-5}$ Pa), and the presence of H$_2$O \citep{polyansky_h2o} and CH$_4$ \citep{exomol_ch4}, using logarithmic uniform priors between $10^{-9}$ and $10^{-2}$. We employed the Multinest \citep{multinest, pymultinest} algorithm with 1500 live points and an evidence tolerance of 0.5 for the fitting procedure. The results of these retrievals are discussed in Sec. \ref{sec:retrieval}.

For each retrieval, we also estimate the Atmospheric Detection Indices (ADI), as defined in \citet{Tsiaras2018}, by comparing the log evidence for an atmospheric model retrieval with a flat retrieval, where we only fit for radius, temperature, and cloud pressure, and considering no trace gasses or molecular features.

\section{Results}
\begin{figure*}
    \centering
    \includegraphics[width=\textwidth,height=\textheight,keepaspectratio]{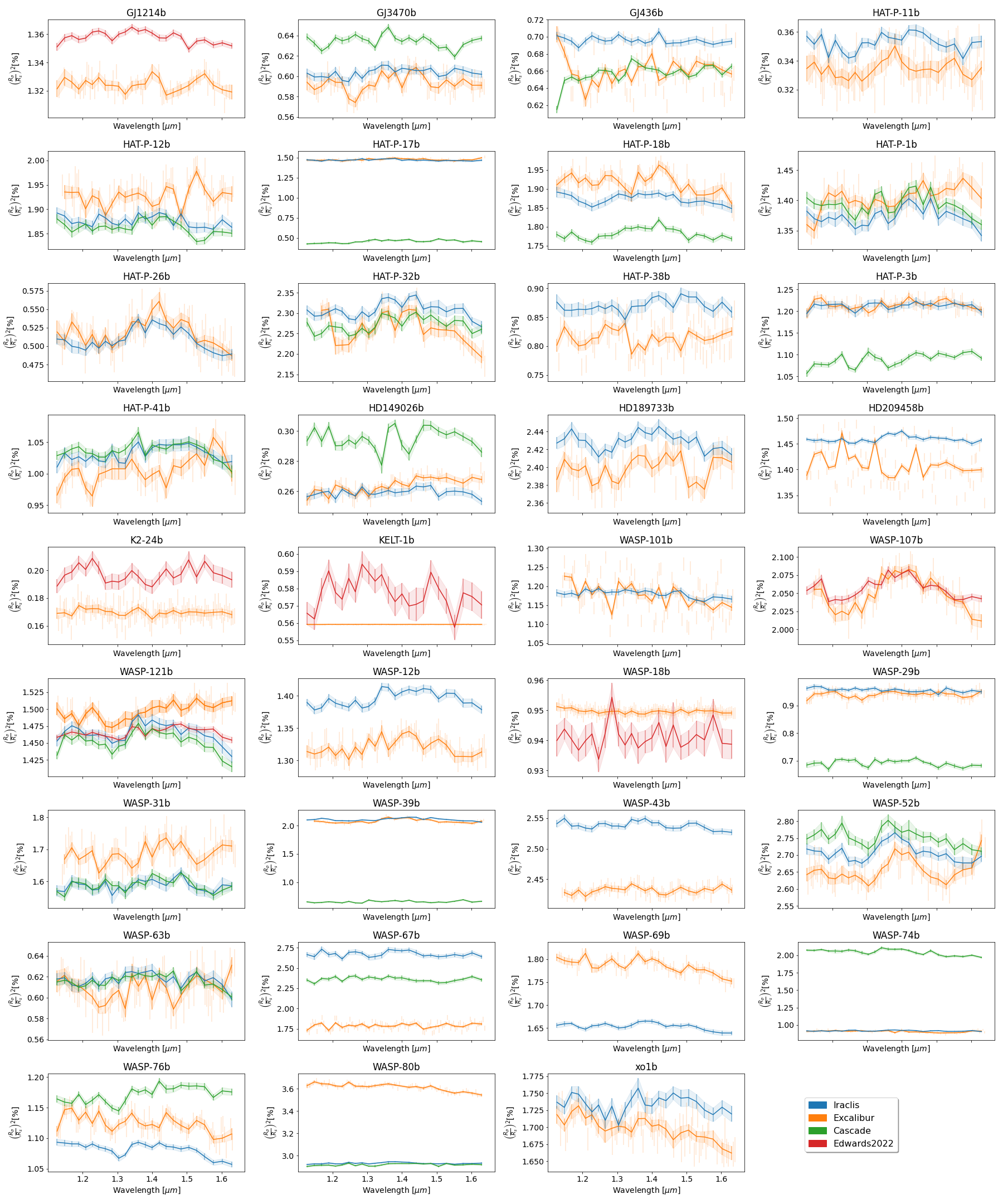}
    \caption{This figure displays all spectra for the 35 planets analyzed in this study. Spectra from the Iraclis dataset are represented in blue, sourced from \citet{Tsiaras2018}. The Edwards2022 dataset spectra, taken from \citet{Edwards2022}, are illustrated in red. In orange, we present the Excalibur dataset spectra, obtained from \citet{Roudier2021}. The unbinned spectra data points are superimposed on the binned one using the same colour. Lastly, spectra produced using the automated CASCADe pipeline for this work are shown in green.}
    \label{fig:spectra}
\end{figure*}

All spectra considered in this study are presented in Fig. \ref{fig:spectra}. Spectra obtained with different pipelines for the same planet are reported on the same panel to ease the comparison. The color code used is consistent with Fig. \ref{fig:database} and with the following figures reported in the manuscript to help the reader in identifying the pipeline used: blue is for Iraclis, red for Edwards2022, orange for Excalibur, and green for Cascade. The y-axis is the panels of Fig. \ref{fig:spectra} is automatically scaled to fit the spectra with their offsets.  

\subsection{Spectra statistics \label{sec:statistics}}
The spectral variations depicted in Fig. \ref{fig:spectra} can be broadly categorized for each planet into three distinct types:
\begin{itemize}
\item Variations in mean values, observable as offsets between the spectra;
\item Discrepancies in claimed precision, manifested as differing error bars for the spectra;
\item Alterations in spectral shape, discernible as varying values in the spectral bins between different spectra, assuming no offset between them.
\end{itemize}
In the subsequent sections, we delve into a detailed analysis of these categories.

\begin{figure*}
\centering
\includegraphics[width=\textwidth]{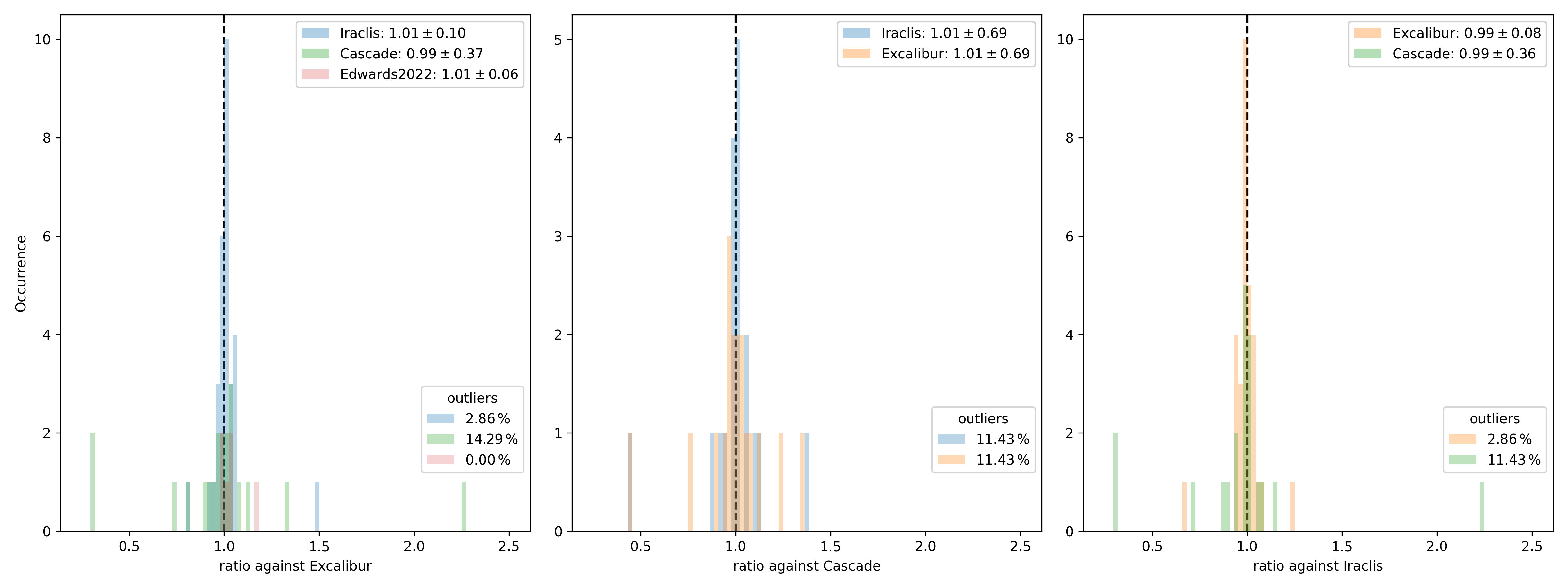}
\caption{This figure displays the ratio of mean spectral values across various exoplanets, defined in the text as $\textrm{MK1}_{A, \, B}$ (eq. \ref{eq:MK1}). Each panel portrays histograms of the mean spectral value ratio. Progressing from left to right, histograms are plotted with respect to the Excalibur dataset, the Cascade dataset, and the Iraclis dataset. The data are divided into 100 uniformly spaced bins ranging from 0 to 2.5. To emphasize the discrepancies in the statistical populations between datasets, occurrences are not normalized. The upper legend denotes the mean values along with standard deviations. The lower legend lists the outliers, which are defined in this context as spectra with means diverging by more than $\pm 25 \%$.}
\label{fig:means}
\end{figure*}

\subsubsection{Variations in mean values \label{sec:mean values}}
Initially, we examine the mean values of the processed spectra, as elaborated in Fig. \ref{fig:spectra}. Fig. \ref{fig:means} demonstrates the ratio of the spectra mean values between different pipelines for the same planet, denoted as $\textrm{MK1}_{A, \, B}$: 
\begin{equation}
\label{eq:MK1}
    \textrm{MK1}_{A, \, B} = \frac{\widehat{Sp}_{A}(\lambda)}{\widehat{Sp}_{B}(\lambda)}
\end{equation}
Here, $Sp(\lambda)$ stands for the measurements in the spectral bin $\lambda$, $A$ and $B$ symbolize different datasets, and $\widehat{Sp}_{A}(\lambda)$ represents the mean value across the spectral bins for that planet in the $A$ dataset.

The left panel of Fig. \ref{fig:means} presents histograms of the ratio of mean values from each pipeline compared to Excalibur. As reported in the legend, median values align closely with one, the expected value for ideal pipelines. Despite the histograms' non-normal distribution, computing the standard deviation aids in the interpretation of these ratios.

While the standard deviation of the Iraclis to Excalibur ratio suggests a ratio of the radii between 0.91 and 1.10, the case for Cascade is more complex. For Cascade, $68\%$ of the ratios lie between 0.62 and 1.36, with a spread that is 3.7 times larger than what was observed in the comparison between Iraclis and Excalibur. This dispersion suggests some pipelines can yield significantly different planet radius estimates. 

The subsequent panels in Fig. \ref{fig:means} adopt Cascade (central panel) and Iraclis (right panel) as references. The right panel echoes the findings of the left, while the central panel reveals similar performance for Excalibur and Iraclis when compared to Cascade. 

We additionally report the number of outliers in the panel, which we define as planets with a ratio of $< 0.75$ or $>1.25$. This definition implies a tolerance of $25\%$ on the ratio of the means. Remarkably, when comparing Cascade and Excalibur, we find that $14.29\%$ of the ratios fall outside of our defined tolerance. The reason for this percentage discrepancy warrants further investigation because such different measurements imply different interpretations of the planetary classes. The discrepancy is highlighted in Fig. \ref{fig:spectra}, with certain planets (e.g., HAT-P-17b, WASP-29b, WASP-39b, WASP-74b) showing significant differences in estimated radii, potentially due to normalization differences \citep{Carone2021}. These four spectra, representing $11.43 \%$ of our sample, are marked as outliers in Fig. \ref{fig:means}. Recalculating the central panel's statistics without these spectra yields an average ratio of $1.00 \pm 0.06$ against Iraclis and $1.00 \pm 0.10$ against Excalibur.

In general, the ratio of the estimated radii is well centered around one, meaning that no pipeline has a preferred bias toward bigger or smaller radii. However, the number of significantly discrepant estimations points to certain planets where the automated procedures of the pipelines yield highly varied radius measurements.

\subsubsection{Discrepancies in uncertainty estimates}
In the following paragraph, we compare the uncertainties and the values computed by the pipeline for each spectral bin.  

\begin{figure*}
    \centering
    \includegraphics[width=\textwidth]{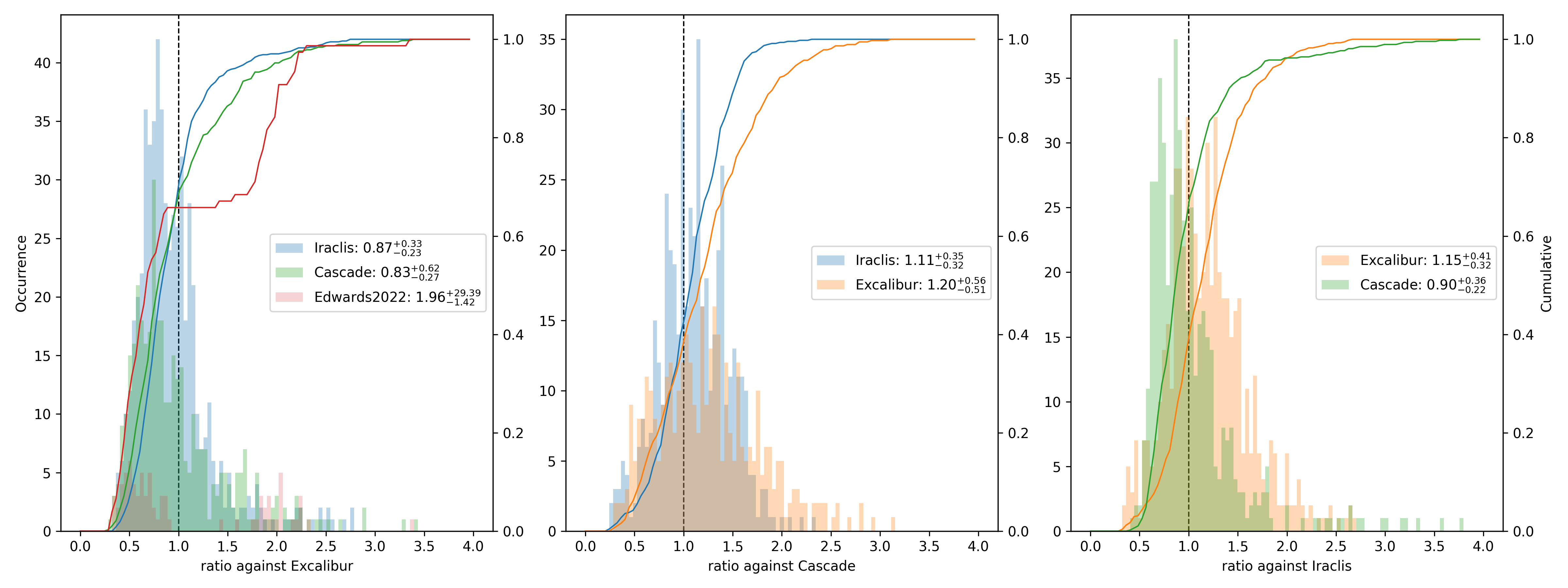}
    \caption{This figure illustrates the ratio between the error bars in identical spectral channels across different planets, defined as $\textrm{MK2}_{A, \, B}$ (eq. \ref{eq:MK2}). Each panel presents histograms of the ratio of the error bars for the same spectral channel. Moving from left to right, histograms are shown with respect to the Excalibur dataset, the Cascade dataset, and the Iraclis dataset. The data are categorized into 100 evenly spaced bins within a range of 0 to 4. Occurrences are not normalized to underline the differences in the statistical populations between datasets. Cumulative curves are superimposed to the histogram and refer to the right y-axis.}
    \label{fig:errorbars}
\end{figure*}

Figure \ref{fig:errorbars} presents a comparative analysis of the error bars across each spectral channel. For each exoplanet, we examine the spectra generated by two distinct pipelines and compare the error bars for each spectral channel, represented as $\textrm{MK2}_{A, \, B}$: 
\begin{equation}
\label{eq:MK2}
    \textrm{MK2}_{A, \, B} =  \frac{\sigma_A(\lambda)}{\sigma_B(\lambda)}   
\end{equation}
where $A$ and $B$ denote the error bars derived from the two pipelines. In the three panels, we use a different pipeline as a reference for each comparison and we report the comparison as histograms with their cumulative curve reported on the right y-axis.

The expectation for pipelines that are consistent with each other is to yield consistent uncertainties across the spectral channels. However, as observed from the first panel, the distribution produced by the six planets from Edwards2022 \citep{Edwards2022} does not align with the data generated by the other pipelines. This discrepancy suggests a potential inconsistency in the pipeline used in Edwards2022 or a unique characteristic of the planets used from that work. However, this discrepancy could also be caused by small number statistics. It is also worth mentioning the case of KELT-1b, where the mean ratio between the uncertainties estimated by Edwards2022 and Excalibur is 34, which suggests pipeline-to-pipeline differences in the interpretation of the planetary system and the data.

Conversely, the Iraclis and Cascade distributions have means at $0.87^{+0.33}_{-0.23}$ and $0.83^{+0.62}_{-0.27}$ respectively, indicating that the Excalibur data set generally exhibits approximately $\sim 15\%$ bigger uncertainties across the spectral channels. The cumulative curves in the panel help highlight this behaviour by reporting the ratio that reaches the saturation limit (1.0) slower. 
In a similar vein, we observe that Iraclis tends to estimate uncertainties that are approximately $10\%$ bigger than those estimated by Cascade.


\begin{figure*}
    \centering
    \includegraphics[width=\textwidth]{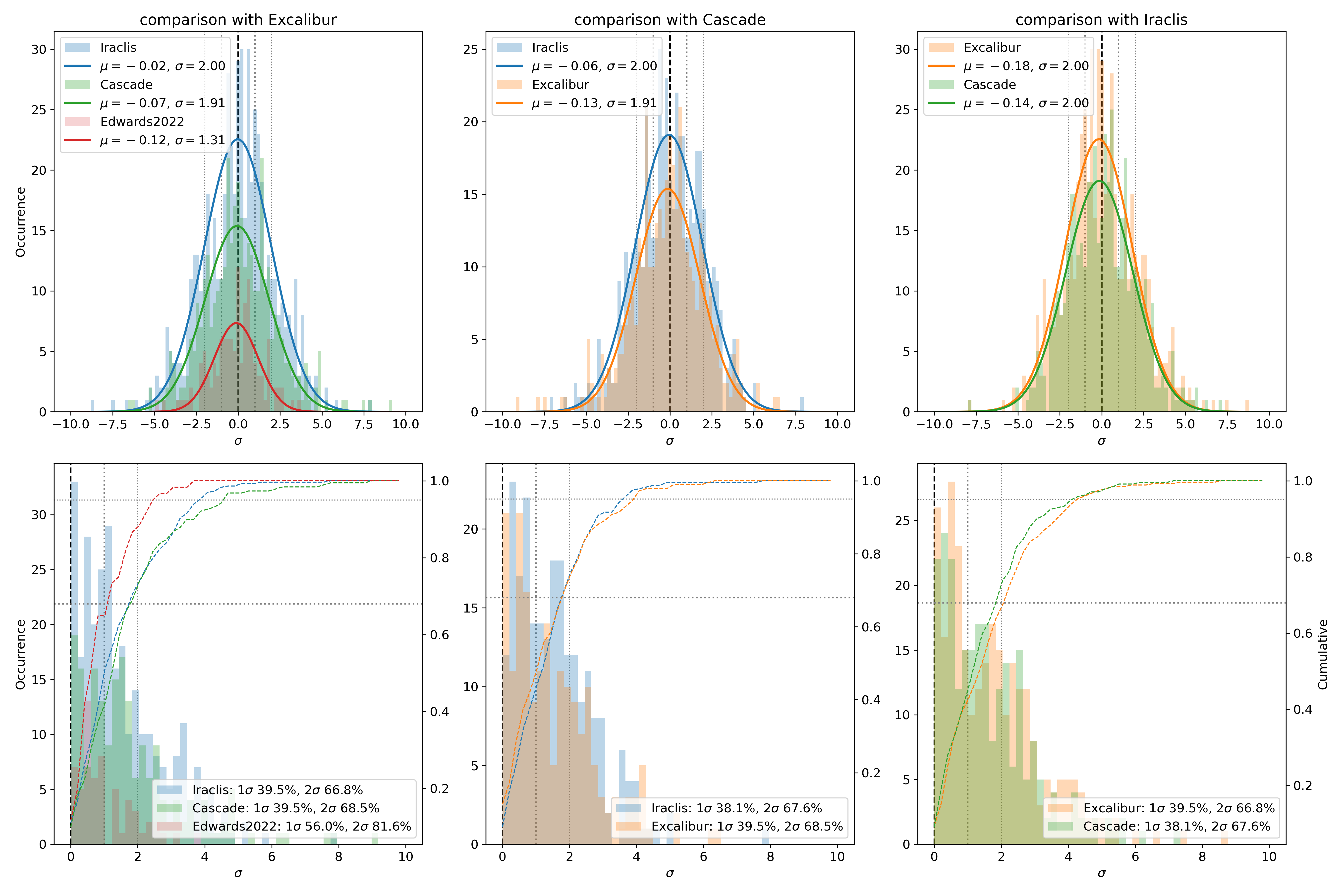}
    \caption{This figure presents histograms of normalized spectral bin differences, where normalization is achieved using the largest standard deviation reported by any pipeline for that specific spectral bin. The top panels show the difference defined as defined in the text as $\textrm{MK3}_{A, \, B}$ (eq. \ref{eq:MK3}). Here, $Sp_{\lambda}$ denotes the values of the spectral bin, and $\sigma_{\lambda}$ represents the standard deviation within that bin. $A$ and $B$ correspond to the two pipelines under comparison. The data are distributed into 100 evenly spaced bins within a range of -10 and 10. To emphasize the distinctions in statistical populations between datasets, occurrences are not normalized. Normal distributions are fitted to the histograms and the fit parameters are reported in the legends.
    The bottom panels show the absolute difference computed as $\textrm{MK4}_{A, \, B}$ (eq. \ref{eq:MK4}), using 50 bins evenly spaced between 0 and 10. The cumulative distribution is overplotted in each panel.
    The legend provides percentages of data points falling within 1 and 2 $\sigma$, also highlighted by the vertical lines in all the panels. The horizontal lines in the bottom panel show the $68\%$ and the $95 \%$ levels desired for $1 \sigma$ and $2 \sigma$ respectively.}
    \label{fig:channels}
\end{figure*}

\subsubsection{Differences in spectral shape}
Figure \ref{fig:channels} illustrates the difference in the estimated values for each spectral channel between two pipelines, normalized by the maximum estimated uncertainties in the channel. To compare the spectra, we also subtract the median values to remove any offset. This is represented as $\textrm{MK3}_{A, \, B}$:
\begin{equation}
\label{eq:MK3}
 \textrm{MK3}_{A, \, B} = \frac{(Sp_{A}(\lambda)-\widehat{Sp}_{A} )-(Sp_{B}(\lambda)-\widehat{Sp}_{B})}{\text{max}[\sigma_{A}(\lambda),\sigma_{B}(\lambda)]}   
\end{equation}
in the top panels and as $\textrm{MK4}_{A, \, B}$:
\begin{equation}
\label{eq:MK4}
\textrm{MK4}_{A, \, B} = \frac{|(Sp_{A}(\lambda)-\widehat{Sp}_{A} )-(Sp_{B}(\lambda)-\widehat{Sp}_{B})|}{\text{max}[\sigma_{A}(\lambda),\sigma_{B}(\lambda)]}
\end{equation}
in the bottom panels, where the absolute value of the difference is considered. $Sp(\lambda)$ signifies the values of the spectral bin, while $\sigma(\lambda)$ denotes the standard deviation within that bin. $A$ and $B$ are placeholders for the two pipelines being compared.

Upon observation, it is evident that there is no significant offset in the distributions. However, the histograms exhibit a broader spread than anticipated. In fact, for consistent spectra, we expect $68 \%$ of the data to be consistent within $1 \sigma$. Specifically, we find that only approximately $40\%$ of the spectral bins yield compatible estimates within the $1 \sigma$ uncertainties. This suggests that the pipelines may not be as consistent with each other as desired, or that the uncertainties are underestimated.
The large distribution, also highlighted by the Normal functions fitted to the data in Fig. \ref{fig:channels} shows no bias, as they are centred around 1, but they also highlight that the spectral features estimated by the pipelines are consistent within $2\sigma$ for around $68 \%$ of the spectral bins. The same estimate is evident from the bottom panels, where are reported the cumulative of the absolute normalized difference. From the horizontal dotted lines set at 0.68 and 0.95, we notice that the cumulative reaches the 0.68 line close to $2 \sigma$, indicated by a dotted vertical line. 

The sole exception to this trend is the comparison between the Excalibur and Edwards2022 pipelines, where compatibility within $1 \sigma$ is achieved for up to $62\%$ of the spectral bins. This higher degree of compatibility could indicate similarities in the methodologies or algorithms used by these two pipelines. However, this comparison might also suffer from the limitations of small number statistics and may not be representative of overall pipeline behaviour. 

It is pertinent to once again highlight the case of KELT-1b. This exoplanet, with a radius of $1.11 \, R_{Jup}$ and a mean density $24 \, \mathrm{g/cm}^3$ \citep{Siverd2012}, exhibits a Transit Spectroscopy Metric (TSM) of 3 as per \citet{kempton2018}, a value that would not predict detectable spectral modulation in its transit spectrum through HST observations. Results from Excalibur \citep{Roudier2021} reveal a flat spectrum, with a mean value of $0.55917 \pm 0.00004$, and the residual standard deviation for individual spectral light curves generally around 2.5 times the photon noise. In contrast, \citet{Edwards2022} identified a more modulated spectrum, with a mean value of $0.577 \pm 0.009$, still compatible with a flat line according to \citet{Edwards2022} Appendix B10, but with clear channel-to-channel correlations in transit depth observed. While analyzing differences in individual targets can be constructive, caution is advised in over-interpreting results based on single-target comparisons. Given the myriad factors that can influence comparisons on individual subjects, we emphasize that pipeline-to-pipeline comparison should ideally be conducted through the analysis of entire catalogues. This approach provides a more robust and replicable framework for evaluating the consistency and reliability of data across different analytical methodologies. Indeed, through systematic comparison across complete catalogues, we can identify trends, consistencies, and discrepancies that are crucial for further refining our observation and analysis techniques, thereby promoting a deeper and more accurate understanding of exoplanetary environments.

\subsection{Retrieval comparison \label{sec:retrieval}}

\begin{figure*}
    \centering
    \includegraphics[width=\textwidth]{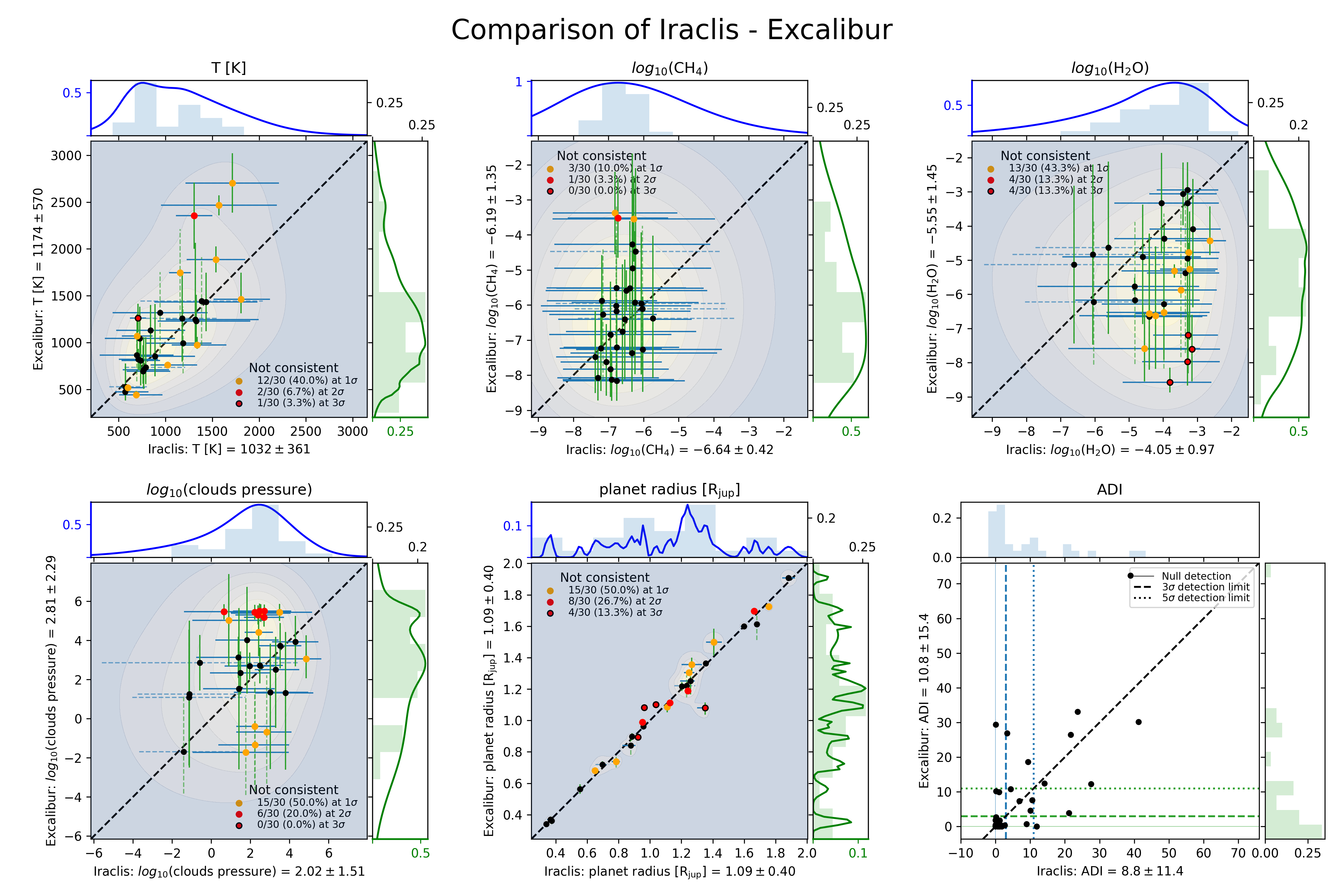}
    \caption{Comparison of retrieval results for Iraclis and Excalibur retrieval results for various planets, marked with blue and green error bars for x and y-axis uncertainties respectively. Compatibility of datasets indicated by alignment with dotted bisector line and color-coded discrepancies: orange for between 1-2 standard deviations ($\sigma$), light red for 2-3$\sigma$, and dark red for more than 3$\sigma$. Parameter space sampling is represented by normalized 2D Normal distributions forming a colored background. Side panels show projected marginal distributions (referring to the colored axis) and histogram references (black axis). The x and y axes labels also report the mean retrieved values with standard deviations. The final panel contrasts Atmospheric Detection Indices (ADIs) with thresholds for 3$\sigma$ (dashed) and 5$\sigma$ (dotted) atmospheric detection.}
    \label{fig:TS_EX}
\end{figure*}

\begin{figure*}
    \centering
    \includegraphics[width=\textwidth]{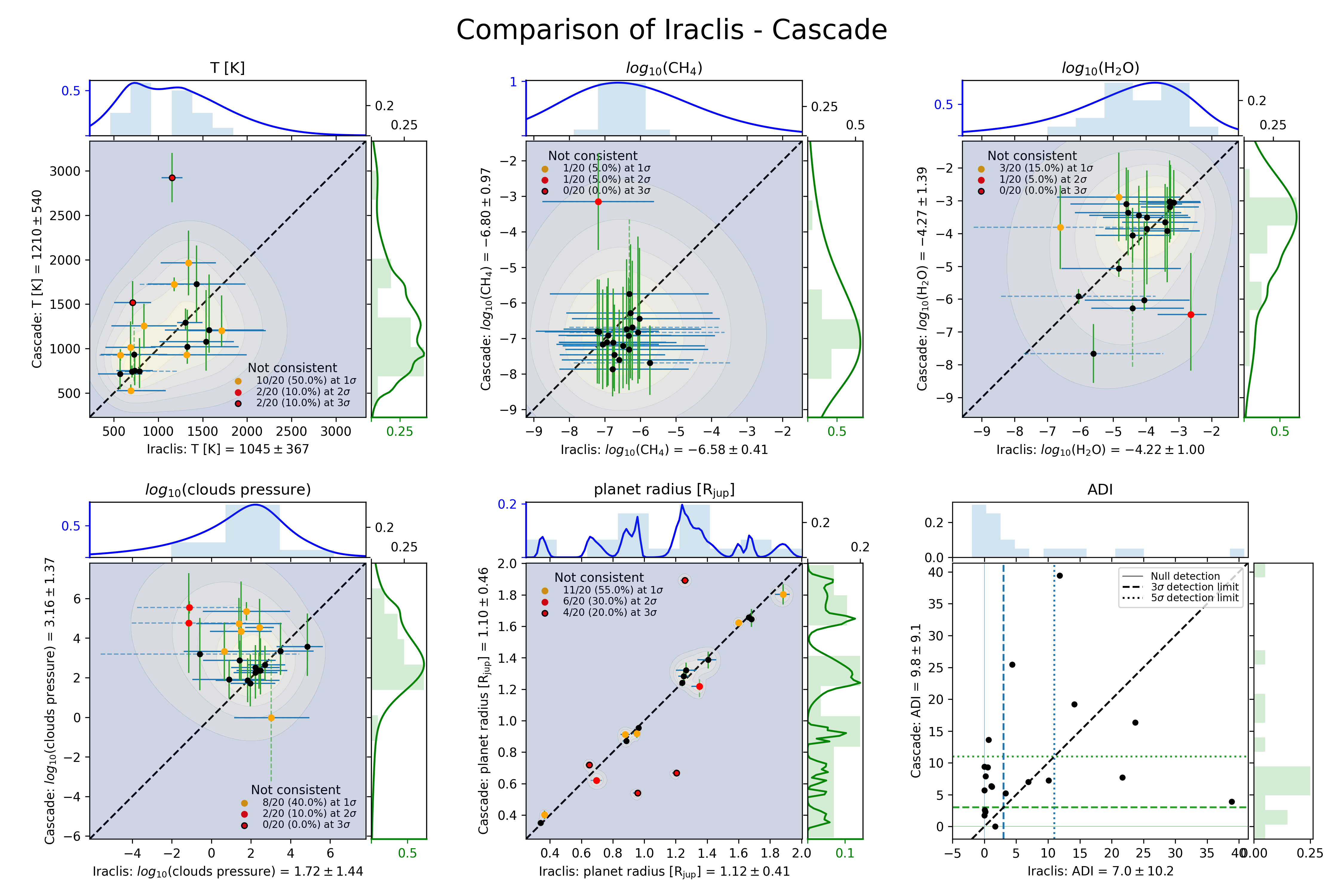}
    \caption{Comparison of retrieval results for Iraclis and Cascade retrieval results for various planets, marked with blue and green error bars for x and y-axis uncertainties respectively. Compatibility of datasets indicated by alignment with dotted bisector line and color-coded discrepancies: orange for between 1-2 standard deviations ($\sigma$), light red for 2-3$\sigma$, and dark red for more than 3$\sigma$. Parameter space sampling is represented by normalized 2D Normal distributions forming a colored background. Side panels show projected marginal distributions (referring to the colored axis) and histogram references (black axis). The x and y axes labels also report the mean retrieved values with standard deviations.The final panel contrasts Atmospheric Detection Indices (ADIs) with thresholds for 3$\sigma$ (dashed) and 5$\sigma$ (dotted) atmospheric detection.}
    \label{fig:TS_CA}
\end{figure*}

To evaluate the discrepancies in the retrieved quantities, we juxtapose the retrieval results for two distinct datasets within the same panel. The results are reported in Fig. \ref{fig:TS_EX}, \ref{fig:TS_CA} and \ref{fig:EX_CA}, where each panel displays the retrieved quantities along with their uncertainties for each planet. Blue error bars correspond to the x-axis, while green error bars are associated with the y-axis. The same color scheme is applied to the side panels (described later in this section). Dotted error bars are used to highlight those retrieved quantities with a null Atmospheric Detection Index (ADI), indicating that a flat line fits the data better than an atmospheric spectrum.

\begin{table*}
    \centering
    \begin{tabular}{|p{15mm}|p{20 mm}|p{20mm}|p{20mm}|p{25mm}|p{20mm}|}
    \hline
         Datasets & \centering T & \centering CH$_4$ & \centering H$_2$O & \centering clouds pressure & planet radius\\
         \hline
         Iraclis \newline Excalibur \newline (Fig. \ref{fig:TS_EX})& 
         WASP-52b & - - & HD209458b \newline HAT-P-41b \newline HAT-P-1b \newline WASP-121b & - -& WASP-80b \newline WASP-69b \newline  WASP-43b \newline  WASP-67b\\
          \hline
        Iraclis \newline Cascade \newline (Fig. \ref{fig:TS_CA}) & 
        WASP-76b \newline WASP-39b & - -& - -& - -& HD149026b \newline WASP-74b \newline WASP-39b \newline HAT-P-17b \\
        \hline
        Excalibur \newline Cascade \newline (Fig. \ref{fig:EX_CA}) &
        HAT-P-41b \newline WASP-121b &- - & HAT-P-41b \newline WASP-63b \newline HAT-P-1b \newline WASP-121b & HAT-P-18b \newline HAT-P-1b & WASP-74b \newline WASP-80b \newline HAT-P-18b \newline WASP-29b \newline WASP-39b \newline HAT-P-17b \\
        \hline
         
    \end{tabular}
    \caption{List of planets not consistent within $3 \sigma$ according to Fig. \ref{fig:TS_EX}, \ref{fig:TS_CA} and \ref{fig:EX_CA}.}
    \label{tab:3sigmas}
\end{table*}

A data point is deemed compatible with the dotted black bisector line if the estimates for the two datasets align. If the bisector falls over $1 \sigma$ from the measurements, the data point is highlighted in orange. If the bisector is over $2 \sigma$, the point is marked in light red, and if it is further than $3 \sigma$, the data point is depicted in red with a black edge. The color-coding scheme used in this analysis provides a visual representation of the level of agreement between the two datasets. It is important to note that the color of a data point does not necessarily reflect the accuracy or reliability of the estimates, but rather the degree of discrepancy between the two datasets. A list of all the planets that are not consistent within $3\sigma$ for each panel and each figure is reported in Tab. \ref{tab:3sigmas}. Upon examining Section \ref{sec:mean values}, we find that HAT-P-17b, WASP-29b, WASP-39b, and WASP-74b were identified as outliers in terms of their spectral mean values. This observation aligns with the expectation that these planets exhibit inconsistencies in their retrieved radius estimations, as highlighted by their discordance within a $3\sigma$ range in Tab. \ref{tab:3sigmas}

To estimate the extent of the parameter space sampled, each data point is replaced with a 2-dimensional Normal distribution, with the uncertainties serving as standard deviations in each direction. These Normal distributions are then normalized to $1$ over the number of data points on the plots and summed. This process generates the colored background of the panels, which is normalized to 1.

The background is subsequently projected in the two directions and displayed in the side panels to represent the marginal distribution. These panels also showcase the histograms for the data point distribution as references.

Lastly, the final panel in Fig. \ref{fig:TS_EX}, \ref{fig:TS_CA} and \ref{fig:EX_CA} presents the comparison between the estimated ADIs. The dashed lines represent the threshold for a $3 \sigma$ detection of an atmosphere, while the dotted line signifies the limit for a $5 \sigma$ detection. The solid line represents a null detection (ADI=0). The lines are color-coded in blue for the horizontal axis and green for the vertical, as the other panels.

\subsubsection{Iraclis and Excalibur}
Fig. \ref{fig:TS_EX} presents a comparison between the retrieval results of Iraclis and Excalibur. The first panel showcases the temperature fit. Only WASP-52b does not have compatible estimates within $3\sigma$, however, $40 \%$ of the planets are not compatible within $1 \sigma$. Overall, the retrieved temperatures between the datasets are generally compatible, but it is noticeable that Excalibur tends to explore the temperature range between $2000 \to 3000 \, K$, while Iraclis spectra do not exceed $2000 \, K$. Additionally, the uncertainties on Excalibur retrieved values are typically smaller than the uncertainties on Iraclis retrieved values.

In terms of the atmospheric features fit, for CH$_4$, Iraclis consistently finds abundances around or below $10^{-6}$, while Excalibur appears to investigate the range $10^{-8} \to 10^{-3}$. Similarly, for water, Iraclis identifies water at about $10^{-5} \to 10^{-3}$ for most of the planets, with some exceptions down to $10^{-6}$. Excalibur uniformly samples the range between $10^{-8}$ (which is a prior dominated region of the parameter space) to $10^{-3}$. There are 4 planets not compatible by more than $3\sigma$, but they are worth noticing because they are in the prior dominated region for Excalibur ($< 10^{-7}$) and in a clear detection area for Iraclis ($> 10^{-4}$). These are HAT-P-1b, HAT-P-41b, HD 209458b, and WASP-121b, as listed in Tab. \ref{tab:3sigmas}. However, it is worth mentioning here that the HST-WFC3 spectra analyzed in this study are all collected in the wavelength range primarily sensitive to water absorption, and they count only 25 data points each in the wavelength range. Therefore, the debate between water and methane can be based only on one or two data points. 

For the cloud pressure panel, it is observed that Iraclis estimates a peak around 316 Pa, while Excalibur's distribution of estimates is biased towards higher values, such as 10000 Pa. Although all planets are compatible within $3\sigma$, there are 6 planets not compatible by more than $2\sigma$ and they all are in the $> 1000$ Pa range for Excalibur estimates.

The subsequent panel reports the planetary radius fit. All these estimates have small error bars, and even though all the values are close to the bisector, there are 8 planets out of 30 that are not compatible by more than $2\sigma$. However, no bias is observed in this panel, as the data points are spread along all the parameter space uniformly.

Finally, the ADI panel indicates that there are some planets for which Excalibur claims a strong detection, while Iraclis does not. This is a reflection of the previous panels which highlights the impact of these differences on our understanding of these exoplanets. The most extreme disagreement however is the single planet for which Iraclis claims a $3 \sigma$ detection, while Excalibur has a null adi: WASP-76b.

\subsubsection{Iraclis and Cascade}
Fig. \ref{fig:TS_CA}, which compares the retrieval results of Iraclis and Cascade, reveals that only $50\%$ of the planets have retrieved temperature estimates that are consistent within $1\sigma$. It is also noticeable that two planets are not consistent within more than $3 \sigma$. One of these is WASP-39b, for which Cascade finds a temperature of $1450  \, K$, while Iraclis finds $763 \, K$. For completeness, Excalibur claims $840 \, K$. The case of WASP-76b is even more striking: Cascade finds $2875 \, K$, while Iraclis finds $1186 \, K$. For comparison, Excalibur estimates $1893 \, K$.

The atmospheric features fit for water and methane are largely consistent between the two datasets, with the notable exception of WASP-39b, which consistently falls between $2 \sigma$ and $3 \sigma$. The discrepancies in retrieved atmospheric content for WASP-39b are correlated with the difference in retrieved temperature. Interestingly, no differences are found in the molecular content for WASP-76b, for which both datasets find no methane but detect water around $10^{-4}$.

The cloud pressure panel in Fig. \ref{fig:TS_CA} reveals discrepancies for 8 planets, 2 of which are not consistent within more than $2 \sigma$. These differences primarily occur in the same area of the parameter space, where Cascade detects clouds from $10^3$ to $10^6$ Pa.

The radius panel exhibits more discrepancies than the others: here, $55 \%$ of the planets have measurements that are not consistent within more than $1 \sigma$, of which 4 planets are not consistent within $3 \sigma$. Contrary to what was observed in Fig. \ref{fig:TS_EX}, here we notice that some measurements are far from the bisector, which may result in different estimates for the planetary class. Two notable examples are WASP-74b and WASP-39b. For WASP-74b, Cascade estimates a radius of $1.89 \, R_{jup}$, while Iraclis estimates $1.24 \, R_{jup}$. For WASP-39b, Cascade estimates $0.67 \, R_{jup}$ and Iraclis estimates $1.20 \, R_{jup}$.

These differences in radius estimates result in discrepancies in the ADI estimates. In fact, we observe in the last panel that for some planets, Iraclis claims a very strong detection for an atmosphere, while Cascade cannot claim any detection, and vice versa. In particular, it is worth noticing that there are two data points for which Iraclis claims a null ADI while Cascade does not: HAT-P-3b, for which  Cascade measures 9.4, and GJ 436b, for which it claims 5.7.

\begin{figure*}
    \centering
    \includegraphics[width=\textwidth]{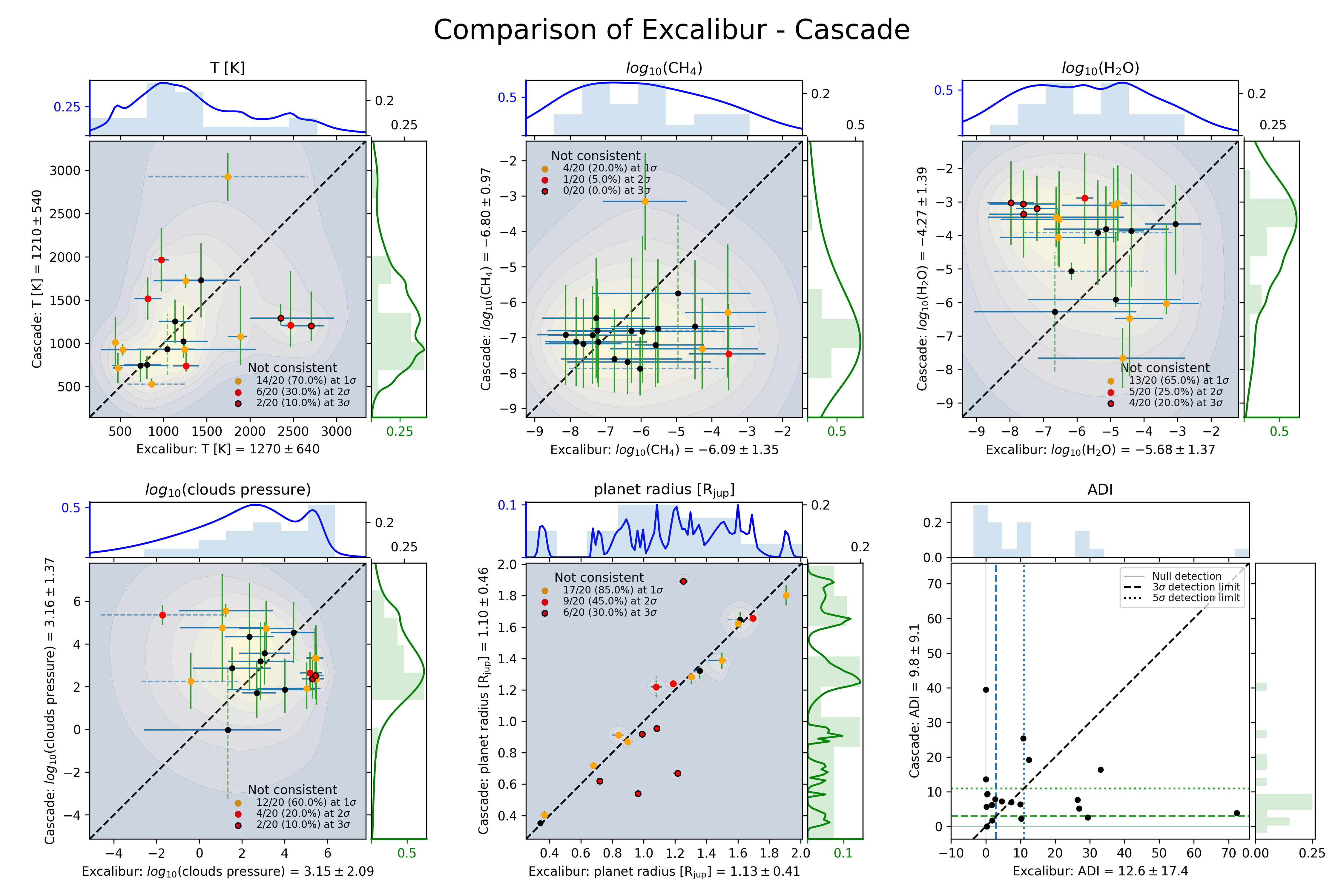}
    \caption{Comparison of retrieval results for Excalibur and Cascade retrieval results for various planets, marked with blue and green error bars for x and y-axis uncertainties respectively. Compatibility of datasets indicated by alignment with dotted bisector line and color-coded discrepancies: orange for between 1-2 standard deviations ($\sigma$), light red for 2-3$\sigma$, and dark red for more than 3$\sigma$. Parameter space sampling is represented by normalized 2D Normal distributions forming a colored background. Side panels show projected marginal distributions (referring to the colored axis) and histogram references (black axis). The x and y axes labels also report the mean retrieved values with standard deviations. The final panel contrasts Atmospheric Detection Indices (ADIs) with thresholds for 3$\sigma$ (dashed) and 5$\sigma$ (dotted) atmospheric detection.}
    \label{fig:EX_CA}
\end{figure*}

\subsubsection{Excalibur and Cascade}
Finally, Fig. \ref{fig:EX_CA} provides a comparison between the Excalibur and Cascade datasets. It should be noted that both pipelines have been run with the same system parameters to extract the spectra. The first panel already reveals that only $30 \%$ of the temperature estimates are consistent within $1\sigma$. There does not appear to be a preferential bias in one of the two datasets, as differences in the estimates up to a factor of 2 are observed along both axes.

In the molecular feature panels, we observe, similarly to Fig. \ref{fig:TS_EX}, that while Excalibur seems more sensitive to methane, Cascade only has estimates between $10^{-6} \to 10^{-8}$. The only exception is WASP-39b, for which Cascade estimates a methane abundance of $10^{-3}$, while Excalibur places it at $10^{-6}$. For comparison, Iraclis claims $10^{-7}$ for this planet.

In the water panel, there is a group of inconsistent measurements in the area of no detection for Excalibur ($10^{-8} \to 10^{-6}$) and strong detection for Cascade ($10^{-4} \to 10^{-3}$). However, there are also three $1\sigma$ discrepancies in the opposite direction, where Cascade claims no water in the atmosphere ($10^{-8} \to 10^{-6}$) and Excalibur claims a large abundance ($10^{-5} \to 10^{-3}$).

The panels for cloud pressure and radius tell a similar story to what was observed in Fig. \ref{fig:TS_CA}. This is expected because, as shown in Fig. \ref{fig:TS_EX}, Excalibur and Iraclis have all the radii measurements close to the bisector.

In the ADI panel, we notice two particularly anomalous data points. These are HAT-P-12b and WASP-76b. For these, Excalibur records a null ADI, while Cascade claims a more than $5\sigma$ detection, with an ADI of 13.6 for HAT-P-12b, and of 39.4 for WASP-76b. For the same planets, Iraclis reports 0.6 and 11.8 respectively.

\section{Discussion}

\subsection{About Edwards2022 dataset}
\label{sec:edwards}
In Sec. \ref{sec:strategy}, we mention our decision to exclude the Edwards2022 dataset from our retrieval analysis. This choice stems from a confluence of reasons, notably the dataset's smaller sample size of six planets and the distribution depicted in Fig. \ref{fig:errorbars}, which deviates from the patterns observed in other datasets. Such a limited dataset does not offer a sufficiently robust basis to deduce population attributes or discern any statistically meaningful deviations from prior Iraclis analyses. 

\subsection{The case of WASP-121b}\label{sec:wasp-121}
In our analysis, WASP-121b emerges as a unique entity, being the sole planet shared across all four datasets. Consequently, we present a comparative study of the spectral retrieval results for all its spectra. It's pertinent to reiterate that both the Iraclis and Edwards2022 datasets employ the Iraclis pipeline for data reduction. Notably, WASP-121b is among the planets previously analyzed in \citet{Tsiaras2018} and subsequently reprocessed in \citet{Edwards2022}. The repeated data detrending is attributed to the data's availability: while \citet{Tsiaras2018} utilized a single transit observation from proposal ID: 14468 by Thomas Mikal-Evans, \citet{Edwards2022} incorporated two transit observations from proposal ID: 15134, again by Thomas Mikal-Evans. As detailed in \citet{Edwards2022}, the analysis extended beyond incorporating two new datasets into the WASP-121b spectrum; it also involved a comprehensive reevaluation of the dataset previously processed in \citet{Tsiaras2018}. This reevaluation was undertaken to ensure uniformity in methodology, parameters, and limb-darkening coefficients across all three transit fits. A comparative review of Tables 8 and 9 from \citet{Edwards2022} with Table 2 from \citet{Tsiaras2018} confirms the consistency in stellar and planetary parameters employed for detrending. Given this uniformity, we posit that the observed differences likely stem from variances in the pipeline (potentially due to an updated version used by \citet{Edwards2022}) or discrepancies in the estimation of limb-darkening coefficients.

Another salient point is that the spectrum in the Cascade dataset was generated using the CASCADe pipeline, incorporating system parameters from \citet{Roudier2021}. This ensures that the initial parameters used to produce the Excalibur dataset spectra are consistent.  

\begin{figure*}
    \centering
    \includegraphics[width=\textwidth]{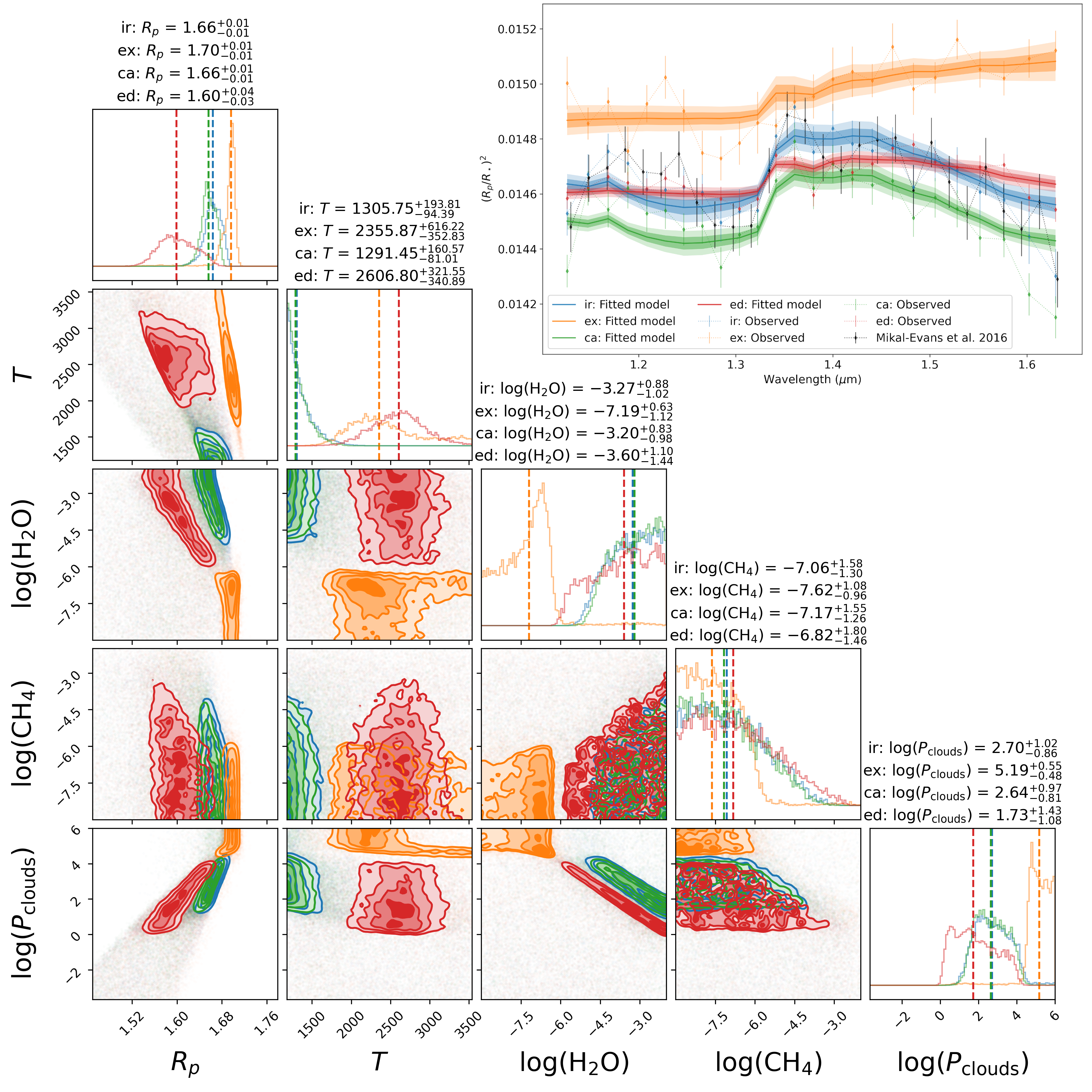}
    \caption{The corner plot delineates the retrieval results for all WASP-121b spectra. The color coding remains consistent with the rest of the manuscript: blue represents the Iraclis dataset, red signifies Edwards2022, orange denotes Excalibur, and green symbolizes Cascade. Atop each panel column, the fitted values for each dataset are displayed: ``ir''corresponds to the Iraclis dataset, ``ex''to Excalibur, ``ca'' to Cascade, and ``ed''to Edwards. The top right panel illustrates the fitted transmission spectra derived from the parameters retrieved for each dataset, represented as solid lines. The corresponding filled areas indicate the $1\sigma$ and $2\sigma$ uncertainties. The observed data points, along with their uncertainties, are depicted using the same color scheme. The observed data points are connected with colored dotted lines, to help the reader. The black data points in the top right panel are from \citet{Evans2016ApJ} for comparison.}
    \label{fig:corner}
\end{figure*}

Upon executing the spectral retrieval as detailed in Sec. \ref{sec:strategy}, we present the resultant corner plots and the derived fitted spectra in Fig. \ref{fig:corner}. The corner plot distinctly demarcates the regions explored by Excalibur (orange) and Edwards (red). In contrast, Iraclis (blue) and Cascade (green) exhibit remarkable congruence across all panels.

The retrieved parameters aren't universally consistent. They yield spectra that pave the way for varied interpretations, especially concerning the planetary temperature, water content, and to a lesser extent, the radius. It's noteworthy that for this planet, Tab. \ref{tab:pln_pars} enumerates an effective temperature (calculated via eq. \ref{eq:temp}) of $2298 \, K$ and a radius of $1.865 \, R_{Jup}$. All pipelines deduce a diminished radius for the planet. Only Edwards2022 and Excalibur yield a temperature in alignment with the projected effective one. In stark contrast, Iraclis and Cascade infer a temperature nearly half the anticipated value, thereby reaching the boundary of our retrieval priors. However, it is known that equilibrium temperatures are often biased to cooler than expected temperatures \citep{MacDonald2020}. On the molecular abundance front, Excalibur uniquely seems to negate the presence of water. Similarly, Excalibur posits that a cloud deck should reside at a pressure higher than that estimated by other pipelines. Broadly, Iraclis and Cascade appear harmonious across every panel, as evidenced by the overlapping blue and green hues. However, a closer inspection of the top-right panel in Fig. \ref{fig:corner} reveals that the blue and green fitted atmospheric modulations aren't congruent within the $2 \sigma$ range. Despite the apparent alignment in retrieved values, the two spectra depict markedly distinct atmospheres.

\section{Conclusions}

In an ideal scenario, all pipelines being compared would be run using the same system parameters and the same exoplanet observations; our analysis is able to make this rigorous, direct comparison for the EXCALIBUR and CASCADe pipelines but, due to the legacy nature of the published Iraclis results, performing a similar analysis was beyond the scope of this initial study. Nonetheless, we believe that including the Iraclis results in the comparison exercise helps illustrate the multifaceted nature of comparing pipelines on a catalogue basis.

Our study shows the significant impact of different data reduction pipelines on the transmission spectra of exoplanets. The analysis of differences in mean values indicates that no pipeline has a bias toward larger or smaller radii compared to the others. However, significant inconsistencies in estimated radius values are common. Regarding the different error bars, we observe that the CASCADe pipeline seems to produce marginally smaller error bars, compared to other pipelines, while EXCALIBUR produces the largest. However, in general, the error bars can be considered consistent within the datasets. The spectral values comparison shows that the shape of the spectra produced with different pipelines seems to be consistent within $2 \sigma$ - however, this approximate consistency is misleading.

From our retrieval analysis, we find there can be significant differences in basic planet parameters such as radius and temperature. In terms of atmospheric composition, we find that Iraclis and Cascade typically yield analogous chemistry results and appear to favor water detection over methane. Conversely, spectra derived from the EXCALIBUR pipeline seem receptive to a broader mixing ratio spectrum for both molecules. It's pivotal to clarify that this study isn't an evaluation of which pipeline offers superior or more trustworthy results. Instead, our objective is to gauge the ramifications of pipeline-induced differences at a population level, paving the way for the comparative planetology era anticipated with telescopes like JWST and Ariel. Further analysis of WASP-121b reveals that, while different pipelines like Excalibur and Cascade can produce varying spectra, consistent atmospheric parameters can still be retrieved from similar spectra, as demonstrated by the CASCADe and Iraclis pipelines. This indicates that despite the large parameter space, comparable spectra can lead to consistent parameter estimation for WASP-121b.

The most significant finding from our study is a clear demonstration that pipelines have the potential to inject systematic bias in the inference of population-level composition trends. The inconsistencies between datasets from different instruments, as well as the potential skewing of analysis due to residual systematics, present challenges that warrant further research.

Reflecting on our initial goals, we confirm that our investigation was aimed at identifying potential systematic biases in the results of exoplanet catalogues, stemming from specific pipelines. This study was not intended to trace the genealogy of pipeline-to-pipeline differences, an analysis that would require a more in-depth comparison of standardized intermediate data products.

In light of our findings, we recommend caution when interpreting results derived from different pipelines; in particular, agreement does not imply an absence of bias. 
We strongly advocate for further research into the systematics that influence the injection of bias and the origin of differences in pipeline results: a comprehensive and methodical examination into the causes of these discrepancies is essential through comparing standardized intermediate data products, in addition to the final transit spectra. Currently, the EXCALIBUR pipeline stands out as the only one to include such standardized intermediate data products in its transmission spectra catalogue, an approach we deem vital for further refining our observation and analysis techniques. Such efforts will not only clarify the sources of biases but also enhance the reliability of pipeline outputs, making them indispensable tools for advancing our understanding of exoplanetary atmospheres. It is through this meticulous scrutiny and resolution of pipeline-induced biases that we can achieve a robust and scientifically sound foundation for exoplanetary characterization.

In the future, we hope that this work will contribute to the development of more consistent and reliable practices in the field of exoplanetary science, ultimately leading to a more accurate understanding of exoplanets and their atmospheres. In particular, we encourage the development of pipelines to perform similar studies to compare the pipeline results not only for single planets but for entire populations, to make use of the statistics to extrapolate some pipeline properties and limitations.  

\section*{Acknowledgements}
The authors thank Dr. Angelos Tsiaras and Dr. Jeroen Bouwman for the useful discussions and collaboration for the development of this paper. The authors also thank Dr. Billy Edwards for sharing the processed spectra produced in \citet{Edwards2022}.
LVM was supported by ASI grant n. 2021.5.HH.O and UKSA grant n. ST/W002507/1.

\section*{Data Availability}

 This work is based upon observations with the NASA/ESA Hubble Space Telescope, obtained at the Space Telescope Science Institute (STScI) operated by AURA, Inc. The publicly available HST observations were obtained from the Hubble Archive which is part of the Mikulski Archive for Space Telescopes (MAST). 



\bibliographystyle{mnras}
\bibliography{main} 








\bsp	
\label{lastpage}
\end{document}